\documentclass[aps,twocolumn,superscriptaddress,english,10pt,nofootinbib,preprintnumbers]{revtex4}

\usepackage{amssymb, amsmath, bm, dcolumn, epsf, graphicx, latexsym, slashed, simplewick,mathtools}
\usepackage[utf8]{inputenc}
\usepackage[normalem]{ulem}

\usepackage{color}

\def\be{\begin{equation}}
\def\ee{\end{equation}}
\def\bea{\begin{eqnarray}}
\def\eea{\end{eqnarray}}
\newcommand{\bsm}[1]

\usepackage{aas_macros}

\usepackage{float}
\usepackage{hyperref}
\usepackage{comment}

\newcommand\calG{{\cal{G}}}
\newcommand\calT{{\cal{T}}}

\usepackage[normalem]{ulem}

\begin{document}

\preprint{MIT-CTP/5806}

\title{Field Equations in Chern-Simons-Gauss-Bonnet Gravity}

\author{Alexis Ortega}
\email{alexis\_ortega@brown.edu}
\affiliation{Department of Physics, Brown University, Providence, RI 02912-1843, USA}
\affiliation{Brown Theoretical Physics Center, Brown University, Providence, RI 02912-1843, USA}

\author{Tatsuya Daniel}
\email{tatsuya\_daniel@brown.edu}
\affiliation{Department of Physics, Brown University, Providence, RI 02912-1843, USA}
\affiliation{Brown Theoretical Physics Center, Brown University, Providence, RI 02912-1843, USA}

\author{Savvas M.~Koushiappas}
\email{savvas\_koushiappas@brown.edu}
\affiliation{Department of Physics, Brown University, Providence, RI 02912-1843, USA}
\affiliation{Brown Theoretical Physics Center, Brown University, Providence, RI 02912-1843, USA}
\affiliation{Center for Theoretical Physics, Massachusetts Institute of Technology, Cambridge, MA 02139, USA}

\begin{abstract}
    We investigate the effects of Chern-Simons-Gauss-Bonnet gravity on fundamental metrics. This theory involves perturbative corrections to general relativity, as well as two scalar fields, the axion and the dilaton, that arise from Chern-Simons and Gauss-Bonnet gravity modifications respectively. The combined Chern-Simons-Gauss-Bonnet gravity is motivated by a wide range of theoretical and phenomenological perspectives, including particle physics, string theory, and parity violation in the gravitational sector. In this work, we provide the complete set of field equations and equations of motion of the Chern-Simons-Gauss-Bonnet modified gravity theory for a suite of fundamental metrics (Friedmann-Lema\^{i}tre-Robertson-Walker, Schwarzschild, spherically symmetric, and perturbed Minkowski), under no prior assumptions on the behavior of the fields. The full set of field equations and equations of motion can be numerically solved and applied to specific observables under certain assumptions, and can be used to place constraints on the Chern-Simons-Gauss-Bonnet modified gravity theory.
\end{abstract}

\maketitle

%%%%%%%%%%%%%%%%%%%%%%%%%%%%%%%%%%%%%%%%%%%%%%%%%%%%%%%%%%%%%%%%%%%%%%%%%%%%%%%%%%%%
%%%%%%%%%%%%%%%%%%%%%%%%%%%%%%%%%%%%%%%%%%%%%%%%%%%%%%%%%%%%%%%%%%%%%%%%%%%%%%%%%%%%
\section{Introduction}
%%%%%%%%%%%%%%%%%%%%%%%%%%%%%%%%%%%%%%%%%%%%%%%%%%%%%%%%%%%%%%%%%%%%%%%%%%%%%%%%%%%%
%%%%%%%%%%%%%%%%%%%%%%%%%%%%%%%%%%%%%%%%%%%%%%%%%%%%%%%%%%%%%%%%%%%%%%%%%%%%%%%%%%%%

Einstein's theory of general relativity (GR) has been shown to agree remarkably well with observations \cite{Hulse1975, Will2006, Antoniadis:2013pzd, Berti:2015itd, LIGO2019-1, LIGO2019-2, EventHorizonTelescope:2019dse, LIGO2021, theligoscientificcollaboration2021tests}.
Despite these successes, there are strong theoretical and experimental arguments that point to the need for extensions to the current formulation of GR. 

From the theoretical point of view, 
challenges toward a quantum theory of gravity suggest that GR may require modifications in the strong field regime \cite{Will2014}. These corrections are generally motivated from a high-energy ultraviolet theory that at low energies leads to corrections to GR as an effective field theory (see e.g.~\cite{Alexander2021}). Moreover, there are a wide range of modified gravity theories and extensions to GR that may be motivated from the fact that GR is non-renormalizable at high energies (for a review on the topic see e.g.,  ~\cite{Capozziello2011, Faraoni2010}). 

From the experimental side, although tests of GR such as the observations of gravitational waves by the LIGO-Virgo-Kagra collaboration have thus far not observed significant deviations, modified gravity theories can have a variety of effects on gravitational waves, which can be characterized by modifications to the gravitational wave amplitude and/or phase. Extensive work has been done to derive the gravitational wave modifications induced by certain modified gravity theories, with the idea that such effects may be detectable by future generation detectors such as the Laser Interferometer Space Antenna (LISA) \cite{LISA:2017pwj} or Einstein Telescope (ET) \cite{Punturo:2010zz}. 

However even though gravitational wave waveforms can be used to test GR, the experimental indication that gravity may not be described exactly by GR comes from cosmology. Even though the Friedmann-Lema\^{i}tre-Robertson-Walker (FLRW) metric is admitted as a solution to the field equations of GR, the presence of a dark energy with a negative equation of state, imposed by a plethora of independent experiments, is rather unsettling and has led to many ideas on how GR can accommodate such behavior on the largest scales  \cite{Tsujikawa2013, Rasanen2011, Clifton2012, Sotiriou2010, DeFelice2010}.

More recently, ever\textcolor{black}{-}increasing precision observations of the expansion rate of the universe \textcolor{black}{have} led to a disagreement between different cosmological observations \cite{2022Abdalla}. Dubbed as the Hubble and $S_8$ tensions, the experimentally obtained value of the present-day expansion of the universe measured locally \cite{2022Pantheon+} seems to be in a disagreement with the value inferred from observations of the Cosmic Microwave Background (CMB) \cite{Planck:2018vyg} under a  $\Lambda$ Cold Dark Matter ($\Lambda \text{CDM}$) cosmological model \cite{Freedman:2021ahq}.  A disagreement (albeit at a lesser significance) is also showcased in amplitude measurements of the variance of fluctuations on scales of $8 \text{h}^{-1} \text{Mpc}$ ($\sigma_8$), or equivalently $S_8$ ($S_8 \equiv \sigma_8 \sqrt{\Omega_m/0.3}$, where $\Omega_m$ is the matter density in the universe) \cite{2021CosmoInterwinedIII}, obtained from the CMB \cite{Planck:2018vyg} and late universe measurements of a wide range of low redshift probes \cite{BOSS2020,2018DESY1,KiDS-1000}.

Solutions to the observed disagreements have been explored extensively in literature \cite{DiValentino:2021izs,PhysRevD.107.083527_SKJS}, from the reshuffling of energy densities at different epochs in the expansion history, such as early dark energy \cite{Poulin:2018cxd}, or changes in the microphysics that governs the universe at small scales, e.g.,  decaying dark matter \cite{Kryiakos2019}. A different method of proposed solutions can be motivated from a change in our understanding of spacetime as used in cosmology, with modifications to GR having also been proposed as alternatives to inflation, dark matter, and dark energy (e.g., \cite{Faraoni2010, Nojiri2007, Nojiri2017, Toomey:2023xsf}). 

In GR, the spacetime manifold is the central object of study; different assumptions when solving the Einstein field equations result in different solutions. If one modifies GR, then the field equations will also be modified, thus changing their solutions. At each position in the manifold, there is a choice to be made besides the coordinate system; one must also choose the values that dictate the geometrical and casual structure of the manifold, i.e. what metric the manifold takes \cite{Schutz2022}. 

The equations of motion provide a prescription to describe past and future events through the geometrical connection between curvature and stress-energy, since energy-momentum from all sources affects the geometry of spacetime. As such, modifications to the field equations can be thought of as additional effects to the geometry through changes to the curvature and stress-energy, describing regions of the manifold where unmodified GR is ill-defined. It is thus appealing to explore the effects of such modifications under various metrics.

One well-studied modification of GR is Chern-Simons gravity\footnote{In almost all instances in this paper, we will be referring to the dynamical formulation of Chern-Simons gravity, known as \textit{dynamical Chern-Simons gravity} \cite{Alexander2009}.}, which can be motivated from the context of particle physics \cite{Alvarez-Gaume1983, Weinberg1996} and leptogenesis \cite{Alexander2006, AlexanderGates2006}, as well as in other areas such as string theory \cite{Green1984, Green:1984ed, Green1987}  and loop quantum gravity \cite{Ashtekar2004, Thiemann2001, Rovelli2004}. Furthermore, from a phenomenological perspective, such a theory could give rise to parity violation in the gravitational sector \cite{Jackiw2003, Contaldi2008, Alexander2009, AlexanderYunes2018, LoutrelYunes2022} and the CMB \cite{QUaD2009, Sorbo2011, Shiraishi2013, Shiraishi2016, Philcox2023}. 

Gauss-Bonnet gravity is another well-motivated modified gravity theory, initially arising from an attempt to generalize GR \cite{Lanczos1932, Lanczos1938, Lovelock1970,  Lovelock1971}; it has also been suggested to arise from string theory \cite{Zwiebach1985, Gross1986, Nepomechie1985, Callan1986, Candelas1985}. Its phenomenological implications have been extensively studied, including its predicted effect on compact objects such as black holes and neutron stars \cite{Moura2006, Guo2008, Maeda2009, Pani2009, Kleihaus2011, Ayzenberg2014, Maselli2015, Kleihaus2016, Kokkotas2017}, and its implications for inflation \cite{Kanti2015, Chakraborty2018, Odintsov2018, Yi2018, Odintsov2019, Rashidi2020}. 

It is well-known that Gauss-Bonnet gravity appears in the 4D gravitational theory predicted by heterotic string theory \cite{Boulware1986, Kanti1996, Torii1997, Alexeyev1997}\footnote{In string theory, the heterotic string is a mixture of the right-moving sector of the superstring and the left-moving sector of the  bosonic string.}, but recently it was shown that Gauss-Bonnet gravity alone cannot be the full theory in four dimensions, as it lacks an axion field. However, an axion field can arise from Chern-Simons gravity; thus, the full 4D gravitational theory coming from the heterotic string is in fact a combination of Chern-Simons and Gauss-Bonnet gravities, a result that does not depend on the choice of compactification \cite{Cano2021}. We refer to this theory as Chern-Simons-Gauss-Bonnet (CS-GB) gravity.

It is imperative to attempt to obtain an experimental signature of any modification to GR. In this paper, we study how field equations are changed in CS-GB gravity, in a set of commonly used and in some ways testable metric inputs. This can be thought of as a first attempt in quantifying ``observational" signatures of CS-GB gravity. In a subsequent paper, we plan to solve the resulting field equations under observational constraints and thus obtain a direct handle at the level to which CS-GB gravity is an allowed modification to GR.

The outline for this paper is as follows: after reviewing the theory and modified field equations in Sec.~\ref{review}, we compute the effects of the modified theory on various metrics in Sec.~\ref{metriceffects}. We discuss directions for future work and conclude in Sec.~\ref{con}.

Throughout this paper, we use geometric units such that $G = c = 1$, and we assume a $(-,+,+,+)$ metric signature; Greek letters ($\mu$,$\nu$,...) range over all spacetime coordinates, Latin letters (i,j,...) range over spatial indices, square brackets denote anti-symmetrization over indices, and derivatives take the form $\partial_\alpha = \frac{\partial}{\partial x^\alpha}$.

\section{Review of CS-GB gravity}\label{review}

We begin with a review of CS-GB gravity and how it can arise from heterotic string theory, a result that was first derived in \cite{Cano2021}. We will use the action given by \cite{Bergshoeff1989}, which is obtained upon supersymmetrization of the Lorentz-Chern-Simons terms.

The starting point is the ten-dimensional heterotic superstring effective action, where we perturbatively add corrections to GR with a parameter $\alpha'$, which is equal to the string tension squared ($\ell_s^2$). The procedure will be as follows: we will first reduce this action to four dimensions by compactifying the extra six dimensions on a torus, and we will rewrite the modified Bianchi identity using a choice of compactification. Introducing the modified Bianchi identity in the action, we will vary the action with respect to one of the dynamical degrees of freedom, and solve the resulting equation by expanding in $\alpha'$ to rewrite the degree of freedom in terms of a Lagrange multiplier. At that point, the action will be in the so-called Jordan frame (equivalently the string frame); we will transform it to the Einstein frame by re-scaling the metric. This will give us the CS-GB action in four dimensions, Eq.~(\ref{eq:action}), from which we will obtain the modified field equations.

\subsection{Derivation of the 4D CS-GB Action}
To begin, the 10D heterotic superstring effective action can be written as
\begin{align}
    \hat{S} &= \frac{g_s^2}{16\pi G_N^{(10)}}\int d^{10}x\sqrt{|\hat{g}|}e^{-2\hat{\phi}}\bigg\{\hat{R} - 4(\partial\hat{\phi})^2 + \frac{1}{12}\hat{H}^2 \nonumber \\ &+ \frac{\alpha'}{8}\hat{R}_{(-)\mu\nu ab}\hat{R}_{(-)}^{\mu\nu ab} + \mathcal{O}(\alpha'^3)\bigg\}, \label{eq:heterotic}
\end{align}
where $\hat{\phi}$ is the dilaton and $\hat{R}_{(-)}$ is the curvature of the torsionful spin connection, $\Omega^a{}_{(-) b}$: 
\begin{align}
    \Omega_{(-)}^a{}_{b} = \omega^a_{~b} - \frac{1}{2}H_{\mu~b}^{~a}dx^{\mu}. \label{eq:spinconnection}
\end{align}
In Eq.~(\ref{eq:spinconnection}), $\omega^a_{~b}$ is the usual spin connection, and $a$ and $b$ are Lorentz indices. $\hat{H}$ is the 3-form field strength associated with the Kalb-Ramond 2-form $\hat{B}$\footnote{In string theory, the Kalb-Ramond field $B_{\mu\nu}$ is a bosonic field which plays a role in the dynamics of the theory; it is analogous to the electromagnetic potential in electromagnetism, and it appears along with the metric tensor and the dilaton to represent a set of massless excitations of the closed string. $H_{\mu\nu\rho}$ is the field-strength tensor associated with the Kalb-Ramond field.},  
\begin{align}
    \hat{H} = d\hat{B} + \frac{\alpha'}{4}\omega_{(-)}^L,
\end{align}
and $\omega_{(-)}^L$ is the Lorentz-Chern-Simons 3-form of the torsionful spin connection. All of the gauge fields are already truncated. 

The asymptotic vacuum expectation value of the dilaton $\langle\hat{\phi}_{\infty}\rangle$ is related to the string coupling constant as $g_s = e^{\langle\hat{\phi}_{\infty}\rangle}$, and $G_N^{(10)} = 8\pi^6g_s^2\ell_s^8$ is the ten-dimensional gravitational constant. $\hat{H}$ satisfies the modified Bianchi identity, 
\begin{align}
    d\hat{H} = \frac{\alpha'}{4}\hat{R}_{(-)~~b}^{~~~~a} \wedge \hat{R}_{(-)~a}^{~~b}, \label{eq:bianchi}
\end{align}
where $\wedge$ is the wedge or exterior product. 
\newline

Now, we reduce Eq.~(\ref{eq:heterotic}) to four dimensions. The minimal consistent truncation possible is a direct product compactification on a six-torus, $\mathcal{M}_4 \times \mathrm{T}^6$, where the metric takes the form
\begin{align}
    d\hat{s}^2 = d\Bar{s}^2 + dz^idz^i,~i = 1,...,6, \label{eq:compactification}
\end{align}
where $d\Bar{s}^2$ is the 4D metric in the Jordan frame, the six-torus is parametrized by the coordinates $z_i \sim z_i + 2\pi\ell_s$, and all the Kaluza-Klein vectors and scalars are taken to be trivial. One can check that this compactification ansatz solves all of the 10-dimensional  equations of motion once the lower-dimensional ones are satisfied, making this a consistent truncation (see~\cite{Cano2021}). This compactification yields the same theory as Eq.~(\ref{eq:heterotic}), except in four dimensions and with a gravitational constant $G_N^{(4)} = G_N^{(10)} / 2\pi\ell_s^6$. 

Using Eq.~(\ref{eq:compactification}), Eq.~(\ref{eq:bianchi}) can be written as
\begin{align}
    \frac{1}{3!}\epsilon^{\mu\nu\rho\sigma}\Bar{\nabla}_{\mu}H_{\nu\rho\sigma} + \frac{\alpha'}{8}\Bar{R}_{(-)\nu\rho\sigma}\Tilde{\Bar{R}}_{(-)}^{\mu\nu\rho\sigma} = 0,
\end{align}
where the dual of the Riemann tensor is given as\footnote{Here $\epsilon^{\mu\nu\alpha\beta}$ is the rank 4 Levi-Civita symbol taken as $\epsilon^{\mu\nu\alpha\beta} = 1/\sqrt{|g|} \Delta^{\mu\nu\alpha\beta}, \, \Delta^{0123} = 1.$ }
\begin{align}
    \Tilde{\Bar{R}}_{(-)}^{\mu\nu\rho\sigma} = \frac{1}{2}\epsilon^{\mu\nu\alpha\beta}\Bar{R}_{(-)\alpha\beta}^{~~~~~~\rho\sigma}.
\end{align}

Introducing the Bianchi identity, Eq.~(\ref{eq:bianchi}), in the action along with a Lagrange multiplier $\varphi$, and integrating Eq.~(\ref{eq:heterotic}) by parts in four dimensions, we obtain
\begin{align}
    \Bar{S} &= \frac{1}{16\pi G_N^{(4)}}\int d^4x\sqrt{|\Bar{g}|}\bigg\{e^{-2(\hat{\phi} - \hat{\phi}_{\infty})}\bigg[\Bar{R} - 4(\partial\hat\phi)^2 \nonumber \\ &+ \frac{1}{12}H^2\bigg] - \frac{1}{3!}H_{\mu\nu\rho}\epsilon^{\mu\nu\rho\sigma}\partial_{\sigma}\varphi + \frac{\alpha'}{8}\mathcal{L}_{R^2} + \mathcal{O}(\alpha'^3)\bigg\}, \label{eq:4dcompact}
\end{align}
where $\mathcal{L}_{R^2}$ consists of the curvature-squared terms in the Lagrangian,
\begin{align}
    \mathcal{L}_{R^2} = e^{-2(\hat{\phi} - \hat{\phi}_{\infty})}\Bar{R}_{(-)\mu\nu\rho\sigma}\Bar{R}_{(-)}^{\mu\nu\rho\sigma} - \varphi\Bar{R}_{(-)\mu\nu\rho\sigma}\Tilde{\Bar{R}}_{(-)}^{{\mu\nu\rho\sigma}}. 
\end{align}
With this, we define $\varphi$ to be the axion. 

To simplify calculating the equations of motion, we vary Eq.~(\ref{eq:4dcompact}) in terms of $H$ to get a relation between $H$ and $\varphi$, thus allowing us to replace $H$ with $\varphi$ in the action. From the variation of $H$, we have that
\begin{align}
    e^{-2(\hat{\phi} - \hat{\phi}_{\infty})}\frac{1}{6}H_{\mu\nu\rho} - \frac{1}{6}\epsilon_{\mu\nu\rho\sigma}\Bar{\nabla}^{\sigma}\varphi + \frac{\alpha'}{8}\frac{\delta\mathcal{L}_{R^2}}{\delta H^{\mu\nu\rho}} = 0. \label{eq:varofh}
\end{align} 
Eq.~(\ref{eq:varofh}) can be solved by doing an expansion in $\alpha'$, i.e. $H = H^{(0)} + \alpha'H^{(1)} + \alpha'^2H^{(2)} + ...$ . This yields, for the first two terms, 
\begin{align}
     H_{\mu\nu\rho}^{(0)}  &= e^{2(\hat{\phi} - \hat{\phi}_{\infty})}\epsilon_{\mu\nu\rho\sigma}\nabla^{\sigma}\varphi, \label{eq:h0} \\ 
     H_{\mu\nu\rho}^{(1)} &= -\frac{3}{4}e^{2(\hat{\phi} - \hat{\phi}_{\infty})}\frac{\delta\mathcal{L}_{R^2}}{\delta H^{\mu\nu\rho}}\bigg|_{H^{(0)}} \label{eq:h1}.
\end{align}
We then plug $H(\varphi)$ back into the action and find that Eq.~(\ref{eq:4dcompact}) can be written as
\begin{align}
    \Bar{S} &= \frac{1}{16\pi G_N^{(4)}}\int d^4x\bigg\{e^{-2(\hat{\phi}-\hat{\phi}_{\infty})}\bigg[\Bar{R} - 4(\partial\hat{\phi})^2\bigg] \nonumber \\ &+ \frac{1}{2}e^{2(\hat{\phi}-\hat{\phi}_{\infty})}(\partial\varphi)^2 + \frac{\alpha'}{8}\mathcal{L}_{R^2}\biggr\rvert_{H^{(0)}} + \mathcal{O}(\alpha'^2)\bigg\}\label{eq:jordanaction}.
\end{align}
To evaluate the four-derivative term $\mathcal{L}_{R^2}$, we have to substitute in Eq.~(\ref{eq:h0}), and use the fact that the curvature $\hat{R}_{(-)}$ can be written in terms of $\hat{H}$ as well as the Riemannian curvature $\hat{R}$:
\begin{align}
    \hat{R}_{(-)\mu\nu~\sigma}^{~~~~~~\rho} &= \hat{R}_{\mu\nu~\sigma}^{~~~\rho} - \hat{\nabla}_{[\mu}\hat{H}_{\nu]~~\sigma}^{~~\rho} - \frac{1}{2}\hat{H}_{[\mu|~\alpha}^{~\rho}\hat{H}_{|\nu]~\sigma}^{~~\alpha}.
\end{align}
Evaluation of the four-derivative term yields
\begin{align}
    \mathcal{L}_{R^2}\biggr\rvert_{H^{(0)}} &= e^{-2(\hat{\phi}-\hat{\phi}_{\infty})}\bigg[\Bar{R}_{\mu\nu\rho\sigma}\Bar{R}^{\mu\nu\rho\sigma} + 6\Bar{G}_{\mu\nu}A^{\mu}A^{\nu} + \frac{7}{4}A^4 \nonumber \\ &- 2\Bar{\nabla}_{\mu}A_{\nu}\Bar{\nabla}^{\mu}A^{\nu} - (\Bar{\nabla}_{\mu}A^{\mu})^2\bigg] - \varphi\Bar{R}_{\mu\nu\rho\sigma}\Tilde{\Bar{R}}^{\mu\nu\rho\sigma} \nonumber \\ &+ \text{total derivatives}, \label{eq:4derivativeterm2}
\end{align}
where $A_{\mu} = e^{2(\hat{\phi} - \hat{\phi}_{\infty})}\partial_{\mu}\varphi$ and $\Bar{G}_{\mu\nu}$ is the Einstein tensor\footnote{Here, the bar denotes that the quantity is the Jordan frame.}.

At this point, we note that Eq.~(\ref{eq:jordanaction}) is in the Jordan frame. It is more favorable to have the action in the Einstein frame to compute the modified field equations. To transform Eq.~(\ref{eq:jordanaction}) into the Einstein frame, we need to rescale the metric:
\begin{align}
    \Bar{g}_{\mu\nu} = e^{2(\hat{\phi} - \hat{\phi}_{\infty})}g_{\mu\nu}. \label{eq:conformalrescaling}
\end{align}

Upon transforming the theory from the Jordan frame to the Einstein frame via the conformal rescaling Eq.~(\ref{eq:conformalrescaling}), the effect on the two-derivative terms in the Lagrangian is rather straightforward to compute:
\begin{align}
    \sqrt{|\Bar{g}|}\mathcal{L}_2 = \sqrt{|g|}\bigg[R + 2(\partial\hat{\phi})^2 + \frac{1}{2}e^{4(\hat{\phi} - \hat{\phi}_{\infty})}(\partial\varphi)^2\bigg].
\end{align}
The effect of the conformal rescaling on the four-derivative term $\mathcal{L}_{R^2}$ requires a lengthier calculation; we need to take into account the transformation of the Riemann tensor and the covariant derivative, and integrate Eq.~(\ref{eq:4derivativeterm2}) by parts multiple times. The end result is
\begin{align}
    \sqrt{|\Bar{g}|}&\mathcal{L}_{R^2}\biggr\rvert_{H^{(0)}} = \sqrt{|g|}\bigg\{e^{-2(\hat{\phi} - \hat{\phi}_{\infty}})\bigg[R_{\mu\nu\rho\sigma}R^{\mu\nu\rho\sigma} \nonumber \\ +~&4R^{\mu\nu}(4\partial_{\mu}\hat{\phi}\partial_{\nu}\hat{\phi} + A_{\mu}A_{\nu}) + R[4\nabla^2\hat{\phi} - 4(\partial\hat{\phi})^2 - 3A^2] \nonumber \\ +~&12(\partial\hat{\phi})^4 + 12(\nabla^2\hat{\phi})^2 + \frac{7}{4}A^4 - 12(\partial_{\mu}\hat{\phi}A^{\mu})^2 \nonumber \\ -~&2A^2(\partial\hat{\phi})^2 - 8A^2\nabla^2\hat{\phi} - 16\partial_{\mu}\hat{\phi}A^{\mu}\nabla_{\alpha}A^{\alpha} - 3(\nabla_{\alpha}A^{\alpha})^2\bigg] \nonumber \\ -~&\varphi R_{\mu\nu\rho\sigma}\Tilde{R}^{\mu\nu\rho\sigma}\bigg\} + \text{total derivatives},
\end{align}
which we can rewrite as
\begin{align}
    \sqrt{\overline{\Bar{g}}}\mathcal{L}_{R^2}\biggr\rvert_{H^{(0)}} = \sqrt{\Bar{g}}\bigg[e^{-2(\hat{\phi} - \hat{\phi}_{\infty})}\mathcal{X}_4 - \varphi R_{\mu\nu\rho\sigma}\Tilde{R}^{\mu\nu\rho\sigma} + \mathcal{L}'\bigg],
\end{align}
where $\mathcal{X}_4 = R^2 - 4R_{\mu\nu}R^{\mu\nu} + R_{\mu\nu\rho\sigma}R^{\mu\nu\rho\sigma}$ is the 4D Gauss-Bonnet density, and we have collected the remaining terms in $\mathcal{L}'$.

Now, consider the zeroth order equations of motion
\begin{align}
    \mathcal{E}_{\mu\nu} &= R_{\mu\nu} + 2\partial_{\mu}\hat{\phi}\partial_{\nu}\hat{\phi} + \frac{1}{2}A_{\mu}A_{\nu}, \label{eq:emunueom}\\
    \mathcal{E}_{\hat{\phi}} &= \nabla^2\hat{\phi} - \frac{1}{2}A^2, \label{eq:ephieom} \\
    \mathcal{E}_{\varphi} &= \nabla_{\mu}A^{\mu} + 2\partial_{\mu}\hat{\phi}A^{\mu}. \label{eq:evarphieom}
\end{align}
After some algebra, $\mathcal{L}'$ can be written in terms of Eqs.~(\ref{eq:emunueom})-(\ref{eq:evarphieom}) as follows:
\begin{align}
    \mathcal{L}' &= e^{-2(\hat{\phi} - \hat{\phi}_{\infty})}\bigg\{4\mathcal{E}_{\mu\nu}\mathcal{E}^{\mu\nu} - \mathcal{E}^2 + 12\mathcal{E}_{\hat{\phi}}^2 + 4\mathcal{E}\mathcal{E}_{\hat{\phi}} - 3\mathcal{E}_{\varphi}^2 \nonumber \\ &+ 2\mathcal{E}_{\hat{\phi}}[A^2 - 4(\partial\hat{\phi})^2] - 4\mathcal{E}_{\varphi}\partial_{\mu}\hat{\phi}A^{\mu}\bigg\}. \label{eq:l'}
\end{align}
We see that all the terms in $\mathcal{L}'$ are proportional to the zeroth-order equations of motion, which means if we redefine the fields
\begin{align}
    g_{\mu\nu} &\rightarrow g_{\mu\nu} + \alpha'\Delta_{\mu\nu}, \\ \hat{\phi} &\rightarrow \hat{\phi} + \alpha'\Delta\hat{\phi}, \\ \varphi &\rightarrow \varphi + \alpha'\Delta\varphi,
\end{align}
then we introduce terms linear in $\alpha'$ that are proportional to the zeroth order equations of motion, which we can therefore use to cancel all the terms in Eq.~(\ref{eq:l'}) \cite{Cano2021}.

Thus, introducing the 4D dilaton $\phi = 2(\hat{\phi} - \hat{\phi}_{\infty})$, we end up with a very simple form of the action in four dimensions:
\begin{align}
    S &= \frac{1}{16\pi}\int d^4x\sqrt{|g|}\bigg[R + \frac{1}{2}(\partial\phi)^2 + \frac{1}{2}e^{2\phi}(\partial\varphi)^2 \nonumber \\ &+ \frac{\alpha'}{8}\bigg(e^{-\phi}\mathcal{X}_4 - \varphi R_{\mu\nu\rho\sigma}\Tilde{R}^{\mu\nu\rho\sigma}\bigg) + \mathcal{O}(\alpha'^2)\bigg], \label{eq:action}
\end{align}
where we have set $G_N  = 1$, and $\mathcal{X}_4 = R^2 - 4R_{\mu\nu}R^{\mu\nu} + R_{\mu\nu\rho\sigma}R^{\mu\nu\rho\sigma}$ is the 4D GB density. 

Eq.~(\ref{eq:action}) is the main result we were after -- the 4-dimensional CS-GB action. We see in Eq.~(\ref{eq:action}) that CS-GB gravity is a correction to GR at linear order in $\alpha'$, with a kinetic coupling between the two scalar fields (which we will simply refer to as the axi-dilaton coupling). 

There has been recent progress in achieving moduli stabilization (see e.g. \cite{Anguelova2010, Gukov2004, Cicoli2013, Baumann2014, Bernardo2022} and references therein, as well as \cite{Brandenberger2023, Mcallister2023} for recent reviews); while these results need to be combined with viable string constructions of particle physics, the emergence of the CS and GB terms as corrections to GR in a low-energy effective field theory is a general prediction of string theory \cite{Zwiebach1985, Alexander2009}.

It is also worth noting that it is rather non-trivial that the only higher derivative corrections of Eq.~(\ref{eq:action}) are the CS and GB terms; there are in principle higher derivative terms that could be present in the action. However, Eq.~(\ref{eq:action}) is a general result, and these terms are not neglected by assuming that the scalar fields are of order $\alpha'$; these terms are just simply not present \cite{Cano2021}.

\subsection{CS-GB Field Equations}
In order to test CS-GB gravity, it is more realistic to consider the theory in the presence of matter degrees of freedom. Adding a matter sector to the action Eq.~(\ref{eq:action}), we have
\begin{align}
    S &= \frac{1}{16\pi}\int d^4x\sqrt{|g|}\bigg[R + \frac{1}{2}(\partial\phi)^2 + \frac{1}{2}e^{2\phi}(\partial\varphi)^2 \nonumber \\ &+ \frac{\alpha'}{8}\bigg(e^{-\phi}\mathcal{X}_4 - \varphi R_{\mu\nu\rho\sigma}\Tilde{R}^{\mu\nu\rho\sigma}\bigg) + \mathcal{O}(\alpha'^2)\bigg] \nonumber \\ &+ S_{\text{mat}}\bigg[\chi,e^{\phi}g_{\mu\nu}\bigg], \label{eq:action1}
\end{align}
where $\chi$ are matter fields.

We find the modified field equations by varying Eq.~(\ref{eq:action1}) with respect to the dilaton, axion, and inverse metric, respectively. This yields
\begin{align}
&\nabla^2\phi = e^{2\phi}(\partial\varphi)^2 - \frac{\alpha'}{8}e^{-\phi}\mathcal{X}_4 - 8\pi T^{\text{mat}}, \label{eq:1} \\
&\nabla_{\mu}(e^{2\phi}\nabla^{\mu}\varphi) = -\frac{\alpha'}{8}R_{\mu\nu\rho\sigma}\Tilde{R}^{\mu\nu\rho\sigma}, \label{eq:2} \\
&G_{\mu\nu} + \frac{\alpha'}{8}\bigg(D_{\mu\nu}^{(\phi)} + 2C_{\mu\nu}\bigg) = 8\pi\bigg(T_{\mu\nu}^{(\phi)} + T_{\mu\nu}^{(\varphi)} + T_{\mu\nu}^{\text{mat}}\bigg), \label{eq:3}
\end{align}
where
\begin{align}
    D_{\mu\nu}^{(\phi)} &= (g_{\mu\rho}g_{\nu\sigma} + g_{\mu\sigma}g_{\nu\rho})\epsilon^{0\sigma\lambda\gamma}\nabla_{\kappa}[^{*}R^{\rho\kappa}_{~~\lambda\gamma}(e^{-\phi})'], \\
    C^{\mu\nu} &= (\nabla_{\alpha}\varphi)\epsilon^{\alpha\beta\gamma(\mu}\nabla_{\gamma}R^{\nu)}_{~\beta} + [\nabla_{(\alpha}\nabla_{\beta)}\varphi]^*R^{\beta(\mu\nu)\alpha}, \\
    T_{\mu\nu}^{(\phi)} &= \nabla_{\mu}\phi\nabla_{\nu}\phi - \frac{1}{2}g_{\mu\nu}\bigg(\nabla_{\alpha}\phi\nabla^{\alpha}\phi\bigg), \\
    T_{\mu\nu}^{(\varphi)} &= e^{2\phi}\nabla_{\mu}\varphi\nabla_{\nu}\varphi - \frac{1}{2}g_{\mu\nu}e^{2\phi}\nabla_{\alpha}\varphi\nabla^{\alpha}\varphi, \\
    T_{\mu\nu}^{\text{mat}} &= e^{2\phi}\bigg[(\rho + P)u_{\mu}u_{\nu} + g_{\mu\nu}P\bigg], \\
    T^{\text{mat}} &\equiv g^{\mu\nu}T_{\mu\nu}^{\text{mat}},
\end{align}
with $D_{\mu\nu}^{(\phi)}$ coming from the variation of the GB term, and $C^{\mu\nu}$ is the C-tensor, a 4-dimensional generalization of the 3-dimensional Cotton-York tensor, which arises from the CS modification to GR. $T^{\text{mat}}_{\mu\nu}$ is the standard matter sector stress-energy \textcolor{black}{tensor} assuming a perfect fluid.

Eqs.~(\ref{eq:1})-(\ref{eq:3}) are a combination of the CS and GB field equations, which is to be expected since in this theory, the CS and GB terms appear as a linear combination at first order in $\alpha'$, in addition to the kinetic coupling between the dilaton and the axion. Eq.~(\ref{eq:1}) shows that the dilaton is sourced by the GB term and the axi-dilaton coupling, in Eq.~(\ref{eq:2}) the Pontryagin term sources the axi-dilaton coupling, and in Eq.~(\ref{eq:3}) a linear combination of CS and GB modifies the GR gravitational field equations.

%%%%%%%%%%%%%%%%%%%%%%%%%%%%%%%%%%%%%%%%%%%%
\subsection{Metric Perturbation Equivalence}
%%%%%%%%%%%%%%%%%%%%%%%%%%%%%%%%%%%%%%%%%%%%

In this subsection, we are interested in how the CS-GB modifications to the field equations, Eq.~(\ref{eq:3}), can be interpreted as perturbations to the spacetime metric. The motivation lies in the fact that the CS-GB modifications enter as perturbative corrections to the field equations, and therefore one may interpret the effect as a perturbation to the underlying spacetime metric. 

In principle, the GR field equations
\begin{align}
    G_{\mu\nu} = 8\pi T_{\mu\nu} \label{eq:efe}
\end{align}
can be calculated given a metric $g_{\mu\nu}$. If we perturb Eq.~(\ref{eq:efe}) by defining
\begin{equation} 
\Tilde{G}_{\mu\nu} \equiv G_{\mu\nu}^{(0)} + \epsilon G_{\mu\nu}^{(1)} + ... 
\end{equation} 
and
\begin{equation} 
\Tilde{T}_{\mu\nu} \equiv T_{\mu\nu}^{(0)} + \epsilon T_{\mu\nu}^{(1)} + ...,
\end{equation}
where $G_{\mu\nu}^{(0)}$ and $T_{\mu\nu}^{(0)}$ are the background Einstein and stress-energy tensors, respectively, and $\epsilon$ is a small parameter, we have
\begin{align}
    \Tilde{G}_{\mu\nu} &= 8\pi \Tilde{T}_{\mu\nu} \\
    G_{\mu\nu}^{(0)} + \epsilon G_{\mu\nu}^{(1)} &= 8\pi\bigg(T_{\mu\nu}^{(0)} + \epsilon T_{\mu\nu}^{(1)}\bigg) \\
    \Rightarrow G_{\mu\nu}^{(1)} &= 8\pi T_{\mu\nu}^{(1)},
\end{align}
since $G_{\mu\nu}^{(0)}$ and $T_{\mu\nu}^{(0)}$ are the background quantities satisfying Eq.~(\ref{eq:efe}). In other words, the perturbations to the field equations can be calculated given another metric $h_{\mu\nu}$ that we can think of as a metric perturbation to the original metric $g_{\mu\nu}$.

To apply this argument to the present work, we  consider Einstein's equations in vacuum, which are simply
\begin{align}
    G_{\mu\nu} = 0. \label{eq:efe-gr}
\end{align}
Perturbing Eq.~(\ref{eq:efe-gr}) by sending $g_{\mu\nu} \rightarrow \Tilde{g}_{\mu\nu} = g_{\mu\nu}^{(0)} + h_{\mu\nu}$, with $g_{\mu\nu}^{(0)}$ the background metric and $h_{\mu\nu}$ perturbations to it, the Einstein field equations are modified to linear order as
\begin{align}
    G_{\mu\nu}^{(0)} + \Box h_{\mu\nu} = 0, \label{eq:efe-grpert}
\end{align}
where $G_{\mu\nu}^{(0)}$ is the background Einstein tensor. 

We can compare Eq.~(\ref{eq:efe-grpert}) to the left-hand side of Eq.~(\ref{eq:3}) (which in vacuum is zero)  and in the  limit where scalar field effects of $\phi$ and $\varphi$ are small enough such that they do not backreact onto the metric. Doing so, we can make the identification that
\begin{align}
    \Box h_{\mu\nu} = \frac{\alpha'}{8}\bigg(D_{\mu\nu}^{(\phi)} + 2C_{\mu\nu}\bigg). \label{eq:metricpertefe}
\end{align}
While Eq.~(\ref{eq:metricpertefe}) in this most general form is not particularly enlightening, we provide solutions for specific metrics in the following section.

%%%%%%%%%%%%%%%%%%%%%%%%%%%%%%%%%%%%%%%%%%%%%%%%%
\section{CS-GB in fundamental geometries}\label{metriceffects}
%%%%%%%%%%%%%%%%%%%%%%%%%%%%%%%%%%%%%%%%%%%%%%%%%
A testable prediction of any modified theory of gravity involves solving the resultant field equations. Solutions to the equations can then be used to test the theory or place constraints on the underlying formalism. In this section we derive explicitly the field equations (Eq.~\ref{eq:3}) for a set of commonly used metric descriptions: 
\begin{itemize} 
\item FLRW metric
\item Spherically symmetric metric
\item Schwarzschild metric
\item Perturbed Minkowski metric
\end{itemize}
In doing so, we do not make any approximations or assumptions on the theory itself, or the range of values that variables can take. With this choice, the results presented here can be applied to any future study that involves CS-GB gravity. 

For clarity, all results from this section (field equations, conservation laws, equations of motion and metric perturbation equivalence equations) are explicitly presented in the Appendix. All {\tt{Mathematica}} notebooks used to derive these results are made publicly available and can be downloaded from~\cite{modCSGB-fieldeqns}.

%%%%%%%%%%%%%%%%%%%%%%%%
\subsection{FLRW Metric}
%%%%%%%%%%%%%%%%%%%%%%%%
We begin with the FLRW metric that describes a homogeneous and isotropic universe: 
\begin{align}
    g_{\mu\nu} = \text{diag} \left\{-1,\frac{a(t)}{1-kr^2},a(t)r^2, a(t)r^2\sin\theta\right\}, 
\end{align}
where as usual $a(t)$ is the scale factor, and $k$ describes the curvature of the hypersurface and can take the values of $k=-1$, $k=0$, and $k=+1$ for negative, zero, and positive curvature. 

Solutions to the unmodified Einstein field equations give the Friedmann equations, which are used to describe the expansion history of the universe. Using the continuity equation, we can reduce the Friedmann velocity equation into a form that is only dependent on the energy densities of the universe \cite{Dodelson2020}, with the energy densities coming from a cosmological constant ($\Lambda$), cold dark matter (CDM), baryonic matter and radiation. Using $\Lambda \text{CDM}$ and values given by the CMB \cite{Planck:2018vyg}, one can then infer the cosmological expansion history given by the Hubble parameter $H(a)$. All observations depend on either  $H(a)$, or its integral. 

Here, we consider adding the CS-GB modifications to the field equations and studying the effects on the FLRW metric. GB gravity has been shown to permit cyclic/bouncing universes, as well as allowing for different allocations of energy densities that change the expansion history via $H$ while remaining consistent with observed measurements \cite{Fernandes:2022zrq}. 

CS gravity has been suggested to lead to parity violation in the early universe with observables in the trispectrum \cite{Shiraishi2016,Creque-Sarbinowski:2023wmb}, as well as providing a possible explanation to the baryon asymmetry, giving rise to leptogenesis \cite{Alexander2009}. If a pseudoscalar field is coupled to the Pontryagin term of CS gravity, it can drive inflation and will subsequently contribute to an inflationary gravitational wave background  \cite{Alexander2009}; from the background, a tensor-to-scalar ratio can be derived, which sets the energy scale of inflation.

Therefore, by evaluating the modifications due to the scalar field\textcolor{black}{s} under an FLRW metric, we investigate the combination of the effects induced by the modified gravity theories. All components of the field equations, continuity, equations of motion, and metric perturbation equivalent equations are provided in Appendix~\ref{sec:FLRW}. We only give the scalar fields temporal dependence to uphold the cosmological principle.

As we are evaluating the theory for an FLRW background, we show that this produces additional effects on the expansion history of the universe. Specifically, using $\dot{a}/a = H$, we get nonlinear contributions to the expansion history. Due to the additional energy contributions from the scalar fields, we can no longer assume that there is the standard conservation of the stress-energy tensor. 

\textcolor{black}{This} additional stress-energy contribution is important when considering the discussion in \cite{Cano:2024anc}, where the issue of applying modified gravity theories on cosmological scales is addressed. Taking the local fundamental properties of a modified gravity theory and ``averaging" them on large scales can lead to inconsistencies in the theory's application to cosmology if the contributions to the stress-energy tensor from the scalar fields are not properly accounted for. 

The change to the conservation of the stress-energy tensor follows from CS-GB gravity, since both contributing terms have non-zero divergence ~\cite{Alexander2009,Fernandes:2022zrq}. In the unmodified case, $8\pi \nabla_\mu T^{\mu\nu} = \nabla_\mu G^{\mu\nu}$ (where $\nabla_\mu G^{\mu\nu} = 0$ due to the Bianchi identities). In the modified case, the Bianchi identities no longer hold due to the CS-GB additions, so we get nonzero contributions from the modified Einstein tensor. This leads to a new form of the continuity equation that now needs to be solved simultaneously with the nonlinear modified Friedmann velocity equation in order to get the expansion history of the universe.

As a sanity check, we can verify that we correctly recover the background FLRW cosmology in the absence of the CS-GB modifications \textcolor{black}{(this is especially applicable to late times).} To do this, we \textcolor{black}{consider} the resultant field equations  \textcolor{black}{and write them in a form that separates the additional modified terms from  the FLRW standard cosmology.}

\textcolor{black}{We take the time-time component of the modified Einstein field equations,
\begin{equation} 
{\cal{G}}_{00} = 8 \pi {\cal{T}}_{00}, 
\end{equation} 
with ${\cal{G}}_{00}$ and ${\cal{T}}_{00}$ taken from Appendix~\ref{sec:FLRW}. With $H = \dot{a}/{a}$ we get (a similar form is found in~\cite{Pinto:2024dnm}),
\begin{align}
   H^2 \left[1+\frac{1}{2}\alpha' \dot{\phi}H \right] = \frac{8\pi} {3}\rho \, \left[e^{2\phi} + \frac{1}{2}e^{2\phi} 
   \frac{\dot{\varphi}^2}{\rho}+\frac{1}{2}\frac{\dot{\phi}^2}{\rho} \right]. \label{eqn:FLRWVel}
\end{align}}
\textcolor{black}{We next calculate the space-space components as 
\begin{equation} 
{\cal{G}}_{ii} = 8 \pi {\cal{T}}_{ii}, 
\end{equation} 
with ${\cal{G}}_{ii}$ and ${\cal{T}}_{ii}$ as before taken from Appendix~\ref{sec:FLRW}. The resultant equation is 
\begin{align}
    \frac{\ddot{a}}{a} \left[ 1 + \frac{1}{2}\alpha' \dot{\phi}H\right] + \frac{1}{2}H^2 \left[ 1 + \frac{1}{2} \alpha' \ddot{\phi} \right]&  \nonumber\\
    = -4\pi P e^{2\phi} - 2\pi e^{2\phi} \dot{\varphi}^2 - 2 \pi \dot{\phi}^2. \label{eqn:FLRWAccel}& 
\end{align}
}
\textcolor{black}{Eqs.~(\ref{eqn:FLRWVel}) and (\ref{eqn:FLRWAccel}) are general expressions applicable to any epoch in the history of the universe. Note that the perturbative coupling term ($\alpha'$) is taken to be small, such that we only consider linear terms in the evaluation of the field equations -- more on this below.}

\textcolor{black}{
\noindent The only non-zero term in the continuity equation is
\begin{equation} 
\nabla_0 {\cal{G}}^{00} = 8 \pi \nabla_0 {\cal{T}}^{00},
\end{equation}
which reduces to 
\begin{align}
    &\dot{\rho}e^{2\phi}+3H\bigg( \rho +P \bigg)  e^{2\phi}\nonumber\\
    +&3H\bigg( \rho +P \bigg)\bigg[\frac{\dot{\phi}}{H}\bigg( \frac{2}{3} \frac{e^{2\phi}\rho}{(\rho+P)} +\frac{1}{3} \frac{e^{2\phi}\dot{\varphi}^2}{(\rho+P)} +\frac{1}{3} \frac{\ddot{\phi}}{(\rho+P)}  \bigg)\nonumber\\
    +&\frac{1}{3}\frac{e^{2\phi}\dot{\varphi}}{H} \frac{\ddot{\varphi}}{(\rho+P)} +\frac{e^{2\phi}\dot{\varphi}^2}{(\rho+P)}+ \frac{\dot{\phi}^2}{(\rho+P)} \bigg] \nonumber\\
    -& \frac{3}{16\pi}\alpha' \dot{\phi}H^2 \frac{\ddot{a}}{a} = 0. \label{eqn:FLRWCont}
\end{align}}

\textcolor{black}{In the absence of coupling and scalar field terms ($\alpha', \phi,\varphi \rightarrow 0$), the CS-GB modifications in the field equations reduce to the standard Friedmann and continuity equations. This is what one expects at late times because the time derivatives of the scalar fields are very small compared to the energy densities \cite{Daniel_2024}. This may not be the case in the early universe, and this is where we expect to have deviations from the FLRW standard cosmology. We study these effects in future follow-up work.}

%%%%%%%%%%%%%%%%%%%%%%%%%%%%%%%%%%%%%%%%%
\subsection{Spherically Symmetric metric}
%%%%%%%%%%%%%%%%%%%%%%%%%%%%%%%%%%%%%%%%%

We now consider a spherically symmetric solution that encompasses both black holes and neutron stars. We consider the following metric:
\begin{align}
    g_{\mu\nu} = \text{diag}\left\{-e^{\nu}, e^{\lambda},r^2, r^2\sin{\theta}\right\}, \label{eq:sphsymmetric}
\end{align}
where $\nu$ and $\lambda$ are functions of $t$ and $r$. The metric includes both interior and exterior solutions, with the interior solutions being important for the study of compact stellar remnants. 

By eliminating the time dependence, interior solutions can be acquired from the non-zero Einstein tensor and the properties of the matter fluid via the conservation of the stress-energy tensor. In the case of only radially dependent $\rho$ and $P$, one can derive the Tolman–Oppenheimer–Volkoff (TOV) equations \cite{PhysRev.55.364,PhysRev.55.374}, which describe the hydrostatic equilibrium of a spherically symmetric body under the influence of gravity. Solving the TOV equations leads to solutions for $\rho$, such as a uniform density distribution or Buchdahl's interior solution \cite{Schutz2022}. Similarly, in the vacuum exterior solution limit, one can derive the Schwarzschild solution; the most general solution for a spherically symmetric metric in a vacuum is the Schwarzschild metric -- see Section~\ref{sec:Schwarzschild-main}.

Now, consider adding time-dependent modifications to the spherically symmetric metric. Normally, the temporal dependence leads to a disagreement in Birkhoff’s theorem, but this can be avoided by working in the scalar-tensor framework of modifications. Introducing scalar-dependent stress-energy tensor contributions leads to a persistent non-vacuum solution, known as a ``hairy solution" \cite{Alexander2009}. The modifications themselves lead to an inherently ``vacuum-less" solution because the scalar fields contribute to the stress-energy tensor in a vacuum. In other words, the most general solution for a spherically symmetric metric has both temporal and radial dependence and takes the form of Eq.~(\ref{eq:sphsymmetric}). 

To investigate Eq.~(\ref{eq:sphsymmetric}) under the CS-GB modifications, we take the scalar fields to be dependent on the spherical coordinates $t,r,\theta$ and $\psi$\footnote{The spherical coordinates $t, r, \theta$ and $\psi$ correspond to $\mu = 0, 1, 2$ and $3$, respectively. Here $\theta$ is the polar angle and $\psi$ is the azimuthal angle.}. The time dependence on the scalar field equations of motion becomes even more sensitive to small changes in the spacetime due to the coupled nature of the scalar potentials to the spacetime -- see Appendix~\ref{sec:NeutronStar}.

As expected, the modifications have a nonzero Einstein tensor and stress-energy tensor. To understand the physics governing the modifications, one must numerically solve the derived coupled equations. Since Eq.~(\ref{eq:sphsymmetric}) is the most general case for a spherically symmetric solution, a plethora of avenues can be taken from the equations in Appendix~\ref{sec:NeutronStar}. 

\subsection{Schwarzschild metric}\label{sec:Schwarzschild-main}

In this subsection we consider the special case of the Schwarzschild metric,  a well-studied, static solution to the Einstein field equations that quantifies the exterior effects induced by black holes \cite{Schwarzschild:1916uq}:
\begin{align}
    g_{\mu\nu} = \text{diag}\left\{ -\bigg(1-\frac{r_s}{r}\bigg), \bigg(1-\frac{r_s}{r}\bigg)^{-1}, r^2, r^2\sin{\theta} \right\}. \label{eq:Schwarzschild}
\end{align}

Black holes are of significant interest to study due to the presence of strong gravity;  effects that would otherwise be negligible due to small couplings are magnified.

In vacuum GR, the Schwarzschild metric produces $G_{\mu\nu} = 0$, which is to be expected because the Schwarzschild metric is a static no-hair solution. A zero Einstein tensor implies that the underlying geometry does not evolve from the initial conditions that determine the structure of the manifold. In other words, once we have a set of conditions from the black hole, we do not expect those conditions to change at some future spacetime in vacuum GR. 

Although the geometry remains unchanged, this is not all the information we can glean from the Schwarzschild metric; we can  determine the changes to the geodesic of a test particle by solving for the leading order changes to its azimuthal angle\footnote{We assume the orbit of the photon is in the equatorial plane ($\theta = \frac{\pi}{2}$) but can be generalized to any deflection around the black hole by rotational symmetry. Here $b$ is the impact parameter and is much larger than $r_s$.} \cite{1992grle.book.....S}:
\begin{align}
    \delta \psi \approx \frac{2 r_s}{b}. \label{eq:deflection angle}
\end{align}

Here, we consider CS-GB gravity as a new description of black holes by considering the Schwarzschild metric as a mechanism to study the changes due to the CS-GB modifications on a well-studied fundamental metric.
In GB gravity, the modifications can take the form as changes to the horizon when solving generically for a spherically symmetric solution with no time dependence \cite{Fernandes:2022zrq,Bryant2021}. The shifting of the horizon can then be studied under gravitational lensing, since the new horizon is a byproduct of the changing geometry \cite{2020JCAP...09..030I}. 

In CS gravity, the Schwarzschild metric and its properties depend heavily on the equations of motion of the axion \cite{Alexander2009}. Within the context of CS-GB gravity, the axion acts as a source term for the dilaton, and additional effects can be calculated by perturbing around the Schwarzschild metric (for example, gravitational wave polarization has been investigated in CS gravity -- see~\cite{Yunes:2007ss,Jackiw2003}).

The modifications induce nonzero contributions to the field equations. As a check, we can reproduce the unmodified field equations by taking the coupling constant $\alpha' \rightarrow 0$, as well as taking $\phi$ and $\varphi$ to zero. As mentioned before, nonzero components to the field equations mean that we have a changing geometry from the initial set of conditions.

The attentive reader will notice the temporal dependence in the scalar fields, even though the Schwarzschild metric is a static solution. In the late universe, all time derivatives of the scalar fields will have to be set to zero, since by that time the scalar fields are in a low energy state. In the early universe, however, the time derivatives of the scalar fields are nonzero, as the scalar fields are high in their potentials and evolving. This detail is important when discussing, for example, the formation of primordial black holes in the very early universe \cite{Zeldovich1966, Hawking1971, Carr2016, Sasaki2018}.

All derived field equations, conservation laws, equations of motion and metric perturbation equivalent equations for the Schwarzschild metric are given in Appendix~\ref{sec:Schwarzschild}. We again assume the dilaton and axion are dependent on the spherical coordinates $t,r,\theta$ and $\psi$\footnote{See footnote 6.}.

\subsection{Minkowski Metric}

The Minkowski metric, which describes a flat manifold with no curvature, is another fundamental metric used in quantum field theory, special relativity, and a variety of GR applications:
\begin{align}
    g_{\mu\nu} = \text{diag} \left\{-1,1,1,1\right\}. \label{eq:minkowski}
\end{align}
A universe can be described by the Minkowski metric in the absence of expansion ($a(t) = 1$). Additionally, any set of time slicing such that $\delta a \sim 0$ will also reproduce the Minkowski metric after normalizing the scaling. As such, local patches of the universe under small infinitesimal time intervals, as well as the universe immediately following inflation, are well described by the Minkowski metric. 

If we are investigating the additional effects of GR modifications in areas of low couplings, it would be difficult to find any effect due to CS-GB gravity in areas with no curvature to begin with. It can be shown that the Einstein field equations are trivially zero, along with any additional modifications from CS-GB gravity. 

While the stress-energy tensor and the scalar field equations of motion are nonzero, the result in a vacuum leads to static fields. This is uninteresting, so instead we perturb the Minkowski metric to investigate any curvature effects that can arise from CS-GB gravity in regions of very low curvature. Perturbing around Minkowski has been used to describe the propagation of gravitational waves \cite{1916SPAW.688E}, as well as to add corrections to Newtonian gravity via the post-Newtonian framework \cite{1915SPAW.47.831E,1917KNAB.19.197D,Lorentz1937} and the post-Minkowskian framework \cite{Kerr:1959zlt,PhysRevD.56.826}. 

\subsubsection{Perturbed Minkowski spacetime and geodesics}
With no large scale expansion, we perturb around Eq.~(\ref{eq:minkowski}) to describe any changes due to the CS-GB modifications as 
\begin{align}
    g_{\mu\nu} = \text{diag} \left\{-(1+2\gamma),(1-2\lambda),(1-2\lambda), (1-2\lambda)\right\}, 
\end{align}
where $\gamma$ and $\lambda$ allow for small perturbations to the spacetime as the scalar fields evolve. To fall in line with the cosmological principle, the perturbations are only given time dependence. 

In Appendix~\ref{sec:PertMinkowski} we provide the resulting field equations, continuity equations, and equations of motion for the perturbed Minkowski case in CS-GB gravity. 

Upon perturbing the Minkowski metric, CS-GB gravity has a direct effect on the geodesics of a photon\footnote{Note that the perturbations are sourced by the scalar fields.}. The geodesic equation is defined as
\begin{align}
    \frac{d}{d\xi}\left(\frac{d x^\mu}{d\xi}\right) + \Gamma^\mu_{\alpha \beta } \frac{dx^\alpha}{d\xi} \frac{dx^\beta}{d\xi} = 0, \label{eq:geodesiceq}
\end{align}
where $\xi$ is the affine parameter of the curve. 
We can subsequently parameterize Eq.~(\ref{eq:geodesiceq}) in terms of a time coordinate $t$. This produces a connection between the coordinate system and the time coordinate,
\begin{align}
   \frac{d^2 x^\mu}{dt^2} + \left(\Gamma^\mu_{\alpha \beta} - \Gamma^0_{\alpha \beta} \frac{dx^\mu}{dt} \right) \frac{dx^\alpha}{dt} \frac{dx^\beta}{dt} = 0.
\end{align}

Restricting the photon to only travel in the $x$-direction, we get the final geodesic equation describing the motion of the photon,
\begin{align}
    \ddot{x}(t)+\dot{x}(t) \left(\frac{2 \dot{\lambda}(t)}{2 \lambda(t)-1}+\frac{\dot{\lambda}(t) \dot{x}(t)^2-\dot{\gamma}(t)}{2 \gamma(t) +1}\right) = 0.
\end{align}
Integrating and solving for $\dot{x}(t)$, we can compare these results to the case of an unperturbed Minkowski metric. In the unperturbed case, we get
\begin{align}
    \dot{x}(t) &= c_1.
\end{align}
We then set $c_1=1$, which matches Einstein's postulate that photons travel linearly at a constant speed\textcolor{black}{,} while in a vacuum. However, in the case of a perturbed Minkowski metric, the motion is described with 
\begin{align}
    \dot{x}(t) &= \frac{\sqrt{2 \gamma (t)+1}}{\sqrt{(2 \lambda (t)-1) (2 c_2 \lambda (t)-1-c_2)}}.
\end{align}
If we take $\lambda$ and $\gamma \rightarrow 0$, we can reproduce Einstein's postulate by setting $c_2=0$, arriving to the same solution as the unperturbed case when setting $c_1=1$. Solving the effects on the geodesics due to CS-GB gravity would require numerically solving the evolution of the scalar fields with respect to the perturbations.

%%%%%%%%%%%%%%%%%%%%%%%%%%%%%%%
\section{Conclusion}\label{con}
%%%%%%%%%%%%%%%%%%%%%%%%%%%%%%%
A number of theoretical and observational challenges suggest that GR may require modifications. Such modifications are assumed to be in the strong gravity regime because for a long time the only experimental  information available has been  in the presence of a weak gravitational field;  observationally testing any modified gravity theory has been a challenge. However, in the last decade, the detection of gravitational waves has transformed the universe into an evolving laboratory by enabling the probing of systems in the strong gravitational regime. Nevertheless, it is expected that generally, any modifications to GR must reproduce observations on cosmological scales, while at the same time provide a consistent explanation of physics on small scales.

The particular modified gravity we explored in this paper, CS-GB gravity,  is well motivated from a variety of theoretical and phenomenological perspectives as we outlined in the introduction. As has been  shown,  heterotic string theory generally predicts this form of a gravity theory in four dimensions, regardless of how the extra dimensions are compactified. The exact choice of how one reduces from ten to four dimensions has implications for non-gravitational matter and fields. In fact, one can even consider a warped compactification\footnote{We thank David Kagan for pointing this out.}, which amounts to adding an additional field that couples to gravity; however, this additional coupling can be hidden in a minimally coupled matter Lagrangian. 

In order to test the validity of any modified theory of gravity, one must connect the theory to observations. This is achieved by computing and solving the field equations, since they provide the equations of motion for any test particle. As a first step towards quantifying cosmological observables for CS-GB gravity, we explored four fundamental metric geometries (FLRW, \textcolor{black}{spherically symmetric, Schwarzschild,} and perturbed Minkowski) and derived the modified field equations for each metric. The utility of these derivations is multifold, as one can take any set of field equations and with appropriate initial and boundary conditions, derive the behavior of observables which in turn can be used to constrain the theory. 

Without solving the field equations, it is important to point out few subtleties in the application and potential interpretation of these results. For example, in the case of the FLRW metric, the CS-GB modifications arise as nonlinear corrections to the expansion history of the universe, in addition to the contributions of the scalar fields to the stress-energy tensor. It is important to accurately represent the gravitational energy contribution in the stress-energy tensor, which is often neglected in modified gravity theories; failing to do so leads to inconsistencies in the macroscopic application \cite{Cano:2024anc}. 

Additionally, as another consistency check with cosmology, one needs to preserve the cosmological principle on large scales. When giving the scalar fields only time dependence, the axion plays a secondary role to the dilaton. Furthermore, as has been shown recently in~\cite{Pinto:2024dnm}, cosmological solutions in scalar GB gravity seem to be indistinguishable from a standard $\Lambda\text{CDM}$ cosmology. Here, we are able to reproduce the behavior observed in \cite{Pinto:2024dnm} by setting the axion terms to zero. In doing so, the equations in Appendix~\ref{sec:FLRW} closely match those in \cite{Pinto:2024dnm}.

In the case of spherically symmetric metrics, there are many directions one can take to quantify  observables in CS-GB gravity. One such direction is to investigate neutron stars in CS-GB gravity by using the modified TOV equations to compute the mass, pressure\textcolor{black}{,} and density of a neutron star by requiring spatial and temporal dependencies for the scalar fields. Additional effects on neutron stars can be quantified by e.g. the mass-radius and I-Love-Q relations \cite{Steiner_2013, Ozel2016, Yagi:2013bca}. Such investigations can shed insight on the interior nuclear physics of a neutron star in the context of modified gravity theories. 

In a relatively simpler fashion (Schwarzschild metric), one can study changes to black hole horizons via gravitational lensing, and how the axi-dilaton coupling modifies results that have already been investigated in, for example, scalar GB gravity alone \cite{Fernandes:2022zrq,Bryant2021}. Moreover, one can derive additional effects by considering perturbations about the Schwarzschild metric. 

Finally, we consider small perturbations to the flat spacetime geometry (Minkowski metric) as a mechanism to describe local patches of the universe $(\delta a \sim 0)$ while the scalar fields evolve in time. The most straightforward observation here is the induced non-zero effects on photon geodesics; however, it is difficult to glean any detailed effect without solving the coupled field equations. 

Future work involves numerically solving the modified field equations (collected in the Appendix). In their most general form and without any simplifying assumptions, it is a rather computationally intensive task.  To facilitate this process, one can consider making assumptions to get a working set of simpler differential equations (for example setting derivatives of fields to zero under certain conditions). However, it is only by these solutions to the modified field equations that we can deepen our understanding of how underlying spacetime geometries are modified via the metric perturbation equivalences outlined in this paper. 

%%%%%%%%%%%%%%%%%%%%%%%%%%%%%%%%%%%%%%%%%%%%%
\section{Acknowledgements}
We acknowledge useful conversations with  Heliudson Bernardo and Michael Toomey. We especially would like to thank Stephon Alexander and David Kagan for insightful comments and stimulating discussions. This research is supported in part by the National Space Grant College and Fellowship Program- Opportunities in NASA 80NSSC20M0053. T.D. is supported by the Simons Foundation, Award 896696. S.~M.~K. is  supported by National Science Foundation (NSF) No. PHY-2412666. The modified equations found in the appendices of this paper were computed in {\tt{Mathematica}} using a symbolic algebra package by Barak Shoshany \cite{Shoshany2021_OGRe}. All {\tt{Mathematica}} notebooks can be found in~\cite{modCSGB-fieldeqns}.

\begin{widetext}

\appendix

%%%%%%%%%%%%%%%%%%%%%%%%%%%%%%%%%%%%%
\section{FLRW metric}\label{sec:FLRW}
%%%%%%%%%%%%%%%%%%%%%%%%%%%%%%%%%%%%%

Field equations: 
\onecolumngrid
\begin{align*}
{\cal{G}}_{00} & =  \frac{3 \left(\dot{a}^2+k\right) \left[\alpha ' \dot{a} \dot{\phi}+2 a\right]}{2 a^3} \\
{\cal{G}}_{11} & =  \frac{2 \left\{\ddot{a} \left[\alpha ' \dot{a} \dot{\phi}+2 a\right]+\dot{a}^2+k\right\}+\alpha ' \ddot{\phi} \left(\dot{a}^2+k\right)}{-2(1- k r^2)} \\
{\cal{G}}_{22} & =  - r^2 (k r - 1) {\cal{G}}_{11} \\
{\cal{G}}_{33 } & =  - r^2 \sin^2\theta (k r - 1) {\cal{G}}_{11} \\
{\cal{G}}_{\mu \nu } & = 0 \hspace{0.5cm} {\mathrm{for}} \hspace{0.5cm} \mu \neq \nu.\\
{\cal{T}}_{00} & =  \frac{1}{2} {\cal{F_\rho}} \\
{\cal{T}}_{11} & =  - \frac{1}{2} \frac{a^2 }{ k r^2-1} {\cal{F_P}}\\
{\cal{T}}_{22} & =  \frac{1}{2} a^2 r^2 {\cal{F_P}} \\
{\cal{T}}_{33 }& =  \frac{1}{2} a^2 r^2 \sin^2\theta {\cal{F_P}} \\
{\cal{T}}_{\mu \nu } & = 0 \hspace{0.5cm} {\mathrm{for}} \hspace{0.5cm} \mu \neq \nu.\\
\end{align*}
where for $\xi =  \left\{\rho,P \right\}$,  
\begin{align*} 
{\cal{F}}_\xi &= e^{2 \phi } \left(\dot{\varphi}^2+2 \xi \right)+\dot{\phi}^2 \\
\end{align*} 
Conservation laws and equations of motion: 
\begin{align*}
\nabla^2\phi &= -\frac{3 \dot{a} \dot{\phi}}{a}-\ddot{\phi}\\
e^{2\phi}(\partial\varphi)^2 - \frac{\alpha'}{8}e^{-\phi}\mathcal{X}_4 - 8\pi T^{\text{mat}} &= -\frac{3 \alpha ' e^{-\phi } \ddot{a} \left(\dot{a}^2+k\right)}{a^3}-e^{2 \phi } \left(\dot{\varphi}^2+24 \pi  P-8 \pi  \rho\right)\\
\nabla_{\mu}(e^{2\phi}\nabla^{\mu}\varphi) &= -\frac{e^{2 \phi } \left\{a \ddot{\varphi}+\dot{\varphi} \left(2 a \dot{\phi}+3 \dot{a}\right)\right\}}{a} \\
-\frac{\alpha'}{8}R_{\mu\nu\rho\sigma}\Tilde{R}^{\mu\nu\rho\sigma} &= 0 \\
\nabla_\mu {\calG}^{\mu \nu} &=\left\{\frac{3 \alpha ' \ddot{a} \dot{\phi} \left(\dot{a}^2+k\right)}{2 a^3},0,0,0\right\} \\
\nabla_\mu {\calT}^{\mu\nu} &= \left\{\frac{e^{2 \phi } \left[3  \dot{a}P+a \left(\dot{\varphi} \left[\ddot{\varphi}+\dot{\phi} \dot{\varphi}\right]+\dot{\rho }\right)+\rho  \left(2 a \dot{\phi}+3 \dot{a}\right)+3 \dot{a} \dot{\varphi}^2\right]+\dot{\phi} \left[a \ddot{\phi}+3 \dot{a} \dot{\phi}\right]}{a},0,0,0\right\}\\
\end{align*}
Metric perturbation equivalence:
\begin{align*}
\Box h_{\mu\nu} &= \frac{\alpha'}{8}\bigg(D_{\mu\nu}^{(\phi)} + 2C_{\mu\nu}\bigg) \\
D_{00}^{(\phi)} &= \frac{12 \dot{a} \dot{\phi} \left(\dot{a}^2+k\right)}{a^3} \\
D_{11}^{(\phi)} &= \frac{-4 \ddot{\phi} \left(\dot{a}^2+k\right)-8 \dot{a} \ddot{a} \dot{\phi}}{1-k r^2} \\
D_{22}^{(\phi)} &= r^2(1-kr^2)  D_{11}^{(\phi)} \\
D_{33}^{(\phi)} &= r^2\sin^2\theta(1-kr^2)  D_{11}^{(\phi)} \\
D_{\mu\nu} &= 0, \, \text{for} \, \mu \neq \nu  \\
C_{\mu\nu} &= 0, \, \text{for all } \mu ,\nu  
\end{align*}

\newpage

%%%%%%%%%%%%%%%%%%%%%%%%%%%%%%%%%%%%%%%%%%%%%%%%%%%%%%%%%%%%%%%%%%%%%%%%%%%%%%%%%%%%%%%%%%%%%%%%%%%%%%%%%%%%%%%%%%%%%
\newpage
\section{Spherically symmetric metric}\label{sec:NeutronStar} 
Field equations:
\begin{align*}
 {\cal{G}}_{00} & =\frac{e^{-2 \lambda } }{4 r^3}\left\{e^\nu \partial_1 \lambda  \left[e^{\lambda } \left(4 r^2-\alpha ' \left[\partial_{22} \phi +\csc ^2\theta \partial_{33} \phi -r \partial_1 \phi +\cot \theta \partial_2 \phi \right]\right)-3 r \alpha ' \partial_1 \phi \right]\right\} \\
 &+\frac{e^{-2 \lambda } }{4 r^3}\left\{\left(e^{\lambda }-1\right) r \left[e^{\lambda } \left(\alpha ' \partial_0 \lambda  \partial_0 \phi +4 e^\nu\right)-2 e^\nu \alpha ' \partial_{11} \phi \right]\right\} \\
 {\cal{G}}_{11} & =\frac{1}{4 r^3}\left\{\alpha ' \left[\left(e^{\lambda }-1\right) r e^{-\nu} \left(\partial_0 \nu \partial_0 \phi -2 \partial_{00} \phi \right)+\partial_1 \nu \left(-\partial_{22} \phi +e^{-\lambda } \left(e^{\lambda }-3\right) r \partial_1 \phi +\csc ^2\theta \left(-\partial_{33} \phi \right)-\cot \theta \partial_2 \phi \right)\right]\right\}\\
 &+\frac{1}{4 r^3}\left\{4 r \left(r \partial_1 \nu-e^{\lambda }+1\right)\right\}\\
 {\cal{G}}_{22} & =\frac{1}{8} \left\{\alpha ' e^{-\lambda -\nu} \left[r \partial_0 \lambda  \left(4 \partial_{01} \phi -\partial_1 \nu \partial_0 \phi \right)-e^\nu \left[2 \partial_{11} \nu-\partial_1 \lambda  \partial_1 \nu+\left(\partial_1 \nu\right)^2\right] \left(\csc ^2\theta \partial_{33} \phi +\cot \theta \partial_2 \phi \right) \right] \right\}\\
 &+\frac{1}{8} \left\{\alpha ' e^{-\lambda -\nu} \left[-r \left(\partial_0 \lambda  \left[\partial_0 \lambda +\partial_0 \nu\right]-2 \partial_{00} \lambda \right) \partial_1 \phi +r \partial_1 \lambda  \left(\partial_0 \nu \partial_0 \phi -2 \partial_{00} \phi \right)\right]\right\}\\
 &+\frac{1}{8} \left\{\alpha ' e^{-2 \lambda }  \left[-r\left(2 \partial_{11} \nu-3 \partial_1 \lambda  \partial_1 \nu+\left(\partial_1 \nu\right)^2\right) \partial_1 \phi -2r \partial_1 \nu \partial_{11} \phi \right]\right\}\\
 &+\frac{1}{8} \left\{\alpha ' \left[-e^{-\nu} \left(\partial_0 \lambda  \left(\partial_0 \nu-\partial_0 \lambda \right)-2 \partial_{00} \lambda \right) \left(\csc ^2\theta \partial_{33} \phi +\cot \theta \partial_2 \phi \right)\right]\right\}\\
 &+\frac{1}{8} \left\{2 r \left[e^{-\lambda } \left(2 r \partial_{11} \nu+\left[r \partial_1 \nu+2\right] \left[\partial_1 \nu-\partial_1 \lambda \right]\right)+r e^{-\nu} \left(\partial_0 \lambda  \left[\partial_0 \nu-\partial_0 \lambda \right]-2 \partial_{00} \lambda \right)\right]\right\}\\
 {\cal{G}}_{33} & = \frac{1}{8} \sin ^2\theta e^{-2 \lambda -\nu} \left\{e^{\lambda +\nu} \left[2 r \left(2 r \partial_{11} \nu+\left[r \partial_1 \nu+2\right] \left[\partial_1 \nu-\partial_1 \lambda \right]\right)-\alpha ' \left(2 \partial_{11} \nu-\partial_1 \lambda  \partial_1 \nu+\left(\partial_1 \nu\right)^2\right) \partial_{22} \phi\right]\right\}\\
 &+ \frac{1}{8} \sin ^2\theta e^{-2 \lambda -\nu} \left\{-\alpha ' re^{\lambda }  \left[\partial_0 \lambda \left(\partial_1 \nu \partial_0 \phi -4 \partial_{01} ^2\phi \right)-2 \partial_{00} \lambda  \partial_1 \phi +2 \partial_1 \lambda  \partial_{00} \phi +\left(\partial_0 \lambda \right)^2 \partial_1 \phi +\partial_0 \nu \left(\partial_0 \lambda  \partial_1 \phi -\partial_1 \lambda \partial_0 \phi \right)\right]\right\}\\
 &+ \frac{1}{8} \sin ^2\theta e^{-2 \lambda -\nu} \left\{ e^{2 \lambda } \left(\partial_0 \lambda  \left(\partial_0 \nu-\partial_0 \lambda \right)-2 \partial_{00} \lambda \right) \left[2 r^2-\alpha ' \partial_{22} \phi \right]\right\}\\
 &+\frac{1}{8} \sin ^2\theta e^{-2 \lambda -\nu} \left\{ -\alpha ' r e^\nu \left[\left(2 \partial_{11} \nu-3 \partial_1 \lambda  \partial_1 \nu+\left(\partial_1 \nu\right)^2\right) \partial_1 \phi +2 \partial_1 \nu \partial_{11} \phi \right]\right\} \\
 {\cal{G}}_{01}&=\frac{e^{-\lambda } \alpha ' }{4 r^3}\left\{\left(e^{\lambda }-1\right) r \left[\partial_1 \nu \partial_0 \phi -2 \partial_{01} ^2\phi \right]+\partial_0 \lambda  \left[-e^{\lambda } \left(\partial_{22} \phi +\csc ^2\theta \partial_{33} \phi-r \partial_1 \phi+\cot \theta \partial_2 \phi \right)-3 r \partial_1 \phi \right]\right\}+\frac{\partial_0 \lambda }{r}={\cal{G}}_{10}\\
 {\cal{G}}_{02}&= \frac{1}{8} \alpha ' \left\{-\frac{2 e^{-\lambda } \left[\partial_0 \lambda \left(\partial_2 \phi -r \partial_{12} ^2\phi\right)+r \partial_1 \lambda \partial_{02} ^2\phi \right]}{r^2} \right\}\\
 &+ \frac{1}{8} \alpha ' \left\{-\frac{1}{8 r^3}\csc \theta e^{\nu-\frac{3 (\lambda +\nu)}{2}} \left[\left(r^2 e^\lambda \left(2 \partial_{00} \lambda+\left(\partial_0 \lambda\right)^2-\partial_0 \lambda \partial_0 \nu\right)+4 e^{\nu+\lambda}\right)\left(\partial_3 \varphi -r \partial_{13} ^2\varphi\right)\right]\right\}\\
 &+ \frac{1}{8} \alpha ' \left\{-\frac{1}{8 r^3}\csc \theta e^{\nu-\frac{3 (\lambda +\nu)}{2}} \left[\left(-re^\nu  \left[2 r \partial_{11} \nu+\left(r \partial_1 \nu-2\right) \left(\partial_1 \nu-\partial_1 \lambda\right)\right]-4e^\nu \right) \left(\partial_3 \varphi -r \partial_{13} ^2\varphi\right)\right]\right\}\\
 &+ \frac{1}{8} \alpha ' \left\{-\frac{1}{8 r^3}\csc \theta e^{\nu-\frac{3 (\lambda +\nu)}{2}} \left[\left(r^2 e^{\lambda }\left[-2 r \partial_{001} ^2\lambda +r \partial_0 \lambda \left(\partial_{01} ^2 \nu-2 \partial_{01} ^2\lambda \right)+\partial_0 \nu \left(r \partial_{01} ^2\lambda +\left(3-r \partial_1 \nu\right) \partial_0 \lambda\right)\right]\right)\partial_3 \varphi\right]\right\}\\
 &+ \frac{1}{8} \alpha ' \left\{-\frac{1}{8 r^3}\csc \theta e^{\nu-\frac{3 (\lambda +\nu)}{2}} \left[ \left(r^2e^{\lambda } \left(2 \left(r \partial_1 \nu-3\right) \partial_{00} \lambda+\left(r \partial_1 \nu-3\right) \left(\partial_0 \lambda\right)^2\right)-4 e^{\nu+\lambda}\right)\partial_3 \varphi\right]\right\}\\
  &+ \frac{1}{8} \alpha ' \left\{-\frac{1}{8 r^3}\csc \theta e^{\nu-\frac{3 (\lambda +\nu)}{2}} \left[\left(4e^\nu-re^\nu \left[r\partial_1 \nu  \left(r \left(\partial_{11} \lambda-2 \partial_{11} \nu\right)-r \left(\partial_1 \lambda\right)^2+\partial_1 \lambda\right)+4\partial_1 \nu \right]\right)\partial_3 \varphi\right]\right\}\\
  &+ \frac{1}{8} \alpha ' \left\{-\frac{1}{8 r^3}\csc \theta e^{\nu-\frac{3 (\lambda +\nu)}{2}} \left[\left( -re^\nu  \left[ r \left(3 r \partial_1 \lambda \partial_{11} \nu-2 \left[r \partial_{111} \nu+\partial_{11} \lambda+2 \partial_{11} \nu\right]+2 \left(\partial_1 \lambda\right)^2\right)\right]\right)\partial_3 \varphi\right]\right\}\\
   &+ \frac{1}{8} \alpha ' \left\{-\frac{1}{8 r^3}\csc \theta e^{\nu-\frac{3 (\lambda +\nu)}{2}} \left[\left( -re^\nu \left[r \left(r \partial_1 \lambda-3\right) \left(\partial_1 \nu\right)^2\right]\right) \partial_3 \varphi \right]\right\} = {\cal{G}}_{20}\\
\end{align*}
\begin{align*}
{\cal{G}}_{03}&= \frac{1}{8} \alpha ' \left\{-\frac{2 e^{-\lambda } \left[\partial_0 \lambda \left(\partial_3 \phi -r \partial_{13} ^2\phi \right)+r \partial_1 \lambda \partial_{03} ^2\phi \right]}{r^2}\right\}\\
&+ \frac{1}{8} \alpha ' \left\{ -\frac{1}{8 r^3}\sin \theta e^{\nu -\frac{3 (\lambda +\nu )}{2}} \left[e^{\lambda } \left(r^2 \left[\partial_0 \lambda \left(\partial_0 \nu-\partial_0 \lambda\right)-2 \partial_{00} \lambda\right]-4 e^\nu \right)\left(\partial_2 \varphi -r \partial_{12} ^2\varphi \right)\right]\right\}\\
&+ \frac{1}{8} \alpha ' \left\{ -\frac{1}{8 r^3}\sin \theta e^{\nu -\frac{3 (\lambda +\nu )}{2}} \left[\left(  re^\nu \left[2 r \partial_{11} \nu+\left(r \partial_1 \nu-2\right) \left(\partial_1 \nu-\partial_1 \lambda\right)\right]+4e^\nu\right) \left(\partial_2 \varphi -r \partial_{12} ^2\varphi \right)\right]\right\}\\
&+ \frac{1}{8} \alpha ' \left\{ -\frac{1}{8 r^3}\sin \theta e^{\nu -\frac{3 (\lambda +\nu )}{2}} \left[\left(r^2 e^{\lambda }\left[2 r \partial_{001} ^2\lambda -r \partial_0 \lambda \left(\partial_{01} ^2 \nu-2 \partial_{01} ^2\lambda \right)+\partial_0 \nu \left(\left(r \partial_1 \nu-3\right) \partial_0 \lambda-r \partial_{01} ^2\lambda \right)\right]\right)\partial_2 \varphi\right]\right\}\\
&+ \frac{1}{8} \alpha ' \left\{ -\frac{1}{8 r^3}\sin \theta e^{\nu -\frac{3 (\lambda +\nu )}{2}} \left[\left(r^2 e^{\lambda }\left[-2 \left(r \partial_1 \nu-3\right) \partial_{00} \lambda+\left(3-r \partial_1 \nu\right) \left(\partial_0 \lambda\right)^2\right]+4 e^{\nu+\lambda} \right)\partial_2\varphi\right]\right\}\\
&+ \frac{1}{8} \alpha ' \left\{ -\frac{1}{8 r^3}\sin \theta e^{\nu -\frac{3 (\lambda +\nu )}{2}} \left[\left(re^\nu  \left[r\partial_1 \nu \left(r \left[\partial_{11} \lambda-2 \partial_{11} \nu\right]-r \left(\partial_1 \lambda\right)^2+\partial_1 \lambda\right)+4\partial_1 \nu\right]\right)\partial_2\varphi\right]\right\}\\
&+ \frac{1}{8} \alpha ' \left\{ -\frac{1}{8 r^3}\sin \theta e^{\nu -\frac{3 (\lambda +\nu )}{2}} \left[\left(re^\nu \left[r \left(3 r \partial_1 \lambda \partial_{11} \nu-2 \left[r \partial_{111} \nu +\partial_{11} \lambda+2 \partial_{11} \nu\right]+2 \left(\partial_1 \lambda\right)^2\right)\right]\right)\partial_2\varphi\right]\right\}\\
&+ \frac{1}{8} \alpha ' \left\{ -\frac{1}{8 r^3}\sin \theta e^{\nu -\frac{3 (\lambda +\nu )}{2}} \left[\left(  re^\nu \left[r \left(r \partial_1 \lambda-3\right) \left(\partial_1 \nu\right)^2\right]-4e^\nu\right) \partial_2 \varphi \right]\right\} = {\cal{G}}_{30}  \\
{\cal{G}}_{12 }&=  \frac{1}{64 r^2}\alpha ' \left\{\csc \theta e^{-\frac{\lambda }{2}-\frac{3 \nu }{2}}  \left[ \left(r^2e^{\lambda } \left[2 \partial_{00} \lambda+\left(\partial_0 \lambda\right)^2-\partial_0 \lambda \partial_0 \nu\right]+4 e^{\nu+\lambda}\right)\partial_{03} \varphi \right]\right\}\\
&+\frac{1}{64 r^2}\alpha ' \left\{ \csc \theta e^{-\frac{\lambda }{2}-\frac{3 \nu }{2}} \left[\left(-re^\nu \left[2 r \partial_{11} \nu+\left(r \partial_1 \nu-2\right) \left(\partial_1 \nu-\partial_1 \lambda\right)\right]-4e^\nu\right) \partial_{03} \varphi\right]\right\}\\
&+\frac{1}{64 r^2}\alpha ' \left\{ \csc \theta e^{-\frac{\lambda }{2}-\frac{3 \nu }{2}} \left[\left( re^\nu  \left[-2 r \partial_{011} ^2 \nu+\left(r \partial_1 \lambda-2 r \partial_1 \nu+2\right) \partial_{01} ^2 \nu+\left(r \partial_1 \nu-2\right) \partial_{01} ^2\lambda \right]\right)\partial_3\varphi\right]\right\}\\
&+\frac{1}{64 r^2}\alpha ' \left\{\csc \theta e^{-\frac{\lambda }{2}-\frac{3 \nu }{2}} \left[\left(e^\nu  \left[r\partial_0 \lambda \left(2 r \partial_{11} \nu+\left[r \partial_1 \nu-2\right] \left[\partial_1 \nu-\partial_1 \lambda\right]\right)+4\partial_0 \lambda\right]\right)\partial_3\varphi\right]\right\}\\
&+\frac{1}{64 r^2}\alpha ' \left\{ \csc \theta e^{-\frac{\lambda }{2}-\frac{3 \nu }{2}} \left[\left(e^{\lambda } r^2 \left[2 \partial_{000} \lambda -\partial_0 \lambda \left(\partial_{00} \nu+\partial_0 \nu \left[\partial_0 \lambda-\partial_0 \nu\right]\right)+\partial_{00} \lambda \left(2 \partial_0 \lambda-3 \partial_0 \nu\right)\right]\right) \partial_3 \varphi \right]\right\}\\
&+\frac{1}{64 r^2}\alpha ' \left\{ 16 \left[e^{-\lambda } \partial_1 \nu \left(r \partial_{12} ^2\phi -\partial_2 \phi \right)-r e^{-\nu } \partial_0 \lambda \partial_{02} ^2\phi \right]\right\} = {\cal{G}}_{21} \\
{\cal{G}}_{13 }&=\frac{1}{64 r^2}\alpha ' \left\{\sin \theta e^{-\frac{\lambda }{2}-\frac{3 \nu}{2}} \left[\left(r^2 e^{\lambda} \left[\partial_0 \lambda \left(\partial_0 \nu-\partial_0 \lambda\right)-2 \partial_{00} \lambda\right]-4 e^{\nu+\lambda}\right)\partial_{02} \varphi\right]\right\}\\
&+\frac{1}{64 r^2}\alpha ' \left\{\sin \theta e^{-\frac{\lambda }{2}-\frac{3 \nu}{2}} \left[\left(re^\nu \left[2 r \partial_{11} \nu+\left(r \partial_1 \nu-2\right) \left(\partial_1 \nu-\partial_1 \lambda\right)\right]+4e^\nu\right) \partial_{02} ^2\varphi \right]\right\} \\
&+\frac{1}{64 r^2}\alpha ' \left\{\sin \theta e^{-\frac{\lambda }{2}-\frac{3 \nu}{2}} \left[\left(e^{\lambda } r^2 \left[-2 \partial_{000} \lambda +\partial_0 \lambda \left(\partial_{00} \nu+\partial_0 \nu \left[\partial_0 \lambda-\partial_0 \nu\right]\right)+\partial_{00} \lambda \left(3 \partial_0 \nu-2 \partial_0 \lambda\right)\right]\right)\partial_2\varphi\right]\right\}\\
&+\frac{1}{64 r^2}\alpha ' \left\{\sin \theta e^{-\frac{\lambda }{2}-\frac{3 \nu}{2}} \left[\left(-re^\nu  \left[-2 r \partial_{011} ^2\nu+\left(r \partial_1 \lambda-2 r \partial_1 \nu+2\right) \partial_{01} ^2 \nu+\left(r \partial_1 \nu-2\right) \partial_{01} ^2\lambda \right]\right)\partial_2 \varphi\right]\right\}\\
&+\frac{1}{64 r^2}\alpha ' \left\{\sin \theta e^{-\frac{\lambda }{2}-\frac{3 \nu}{2}} \left[\left(-e^\nu \left[r\partial_0 \lambda \left(2 r \partial_{11} \nu+\left[r \partial_1 \nu-2\right] \left[\partial_1 \nu-\partial_1 \lambda\right]\right)+4\partial_0 \lambda\right]\right) \partial_2 \varphi \right]\right\}\\
&+\frac{1}{64 r^2}\alpha ' \left\{ 16 \left[e^{-\lambda } \partial_1 \nu \left(r \partial_{13} ^2\phi -\partial_3 \phi \right)-r e^{-\nu} \partial_0 \lambda \partial_{03} ^2\phi \right]\right\} = {\cal{G}}_{31} \\
{\cal{G}}_{23 }&=  -\frac{1}{8} \alpha ' e^{-\lambda -\nu} \left\{e^\nu \left[2 \partial_{11} \nu-\partial_{1} \lambda \partial_{1} \nu+\left(\partial_{1} \nu\right)^2\right]+e^{\lambda } \left[\partial_{0} \lambda \left(\partial_0 \nu-\partial_{0} \lambda\right)-2 \partial_{00} \lambda \right]\right\} \left(\cot \theta \partial_3 \phi -\partial_{23} ^2\phi \right) = {\cal{G}}_{32 } \\
& \\
\end{align*}
\begin{align*}
{\cal{T}}_{00} & =  \frac{1}{2r^3} \left(e^{2 \phi}  {\cal{A}}^{\text{NS}}_{\varphi}+{\cal{A}}^{\text{NS}}_{\phi}+2 \rho r^3  e^{2 \phi  +\nu}\right) \\
{\cal{T}}_{11} & =  \frac{e^{\lambda-\nu}}{2r^3} \left(e^{2 \phi}  {\cal{B}}^{\text{NS}}_{\varphi}+{\cal{B}}^{\text{NS}}_{\phi}+2 P r^3  e^{2 \phi +\nu}\right) \\
{\cal{T}}_{22} & =  \frac{e^{-\nu}}{2r} \left(e^{2 \phi}  {\cal{C}}^{\text{NS}}_{\varphi}+{\cal{C}}^{\text{NS}}_{\phi}+2 P r^3  e^{2 \phi  +\nu}\right) \\
{\cal{T}}_{33} & =  \frac{e^{-\nu} \sin^2 \theta}{2r} \left(e^{2 \phi}  {\cal{D}}^{\text{NS}}_{\varphi}+{\cal{D}}^{\text{NS}}_{\phi}+2 P r^3  e^{2 \phi +\nu}\right) \\
{\cal{T}}_{\mu \nu}&={ e^{2 \phi  } \partial_\mu \varphi  \partial_\nu \varphi +\partial_\mu \phi  \partial_\nu \phi = \cal{T}}_{\nu \mu} , \, \, \,  \mu \neq \nu \\
\end{align*}
where for $\omega =  \left\{ \varphi,\phi \right\}$, 
\begin{align*} 
{\cal{A}}_\omega &=r^3 \left( \partial_0 \omega \right)^2 +  r e^\nu \left( r^2 e^{-\lambda}\left( \partial_1 \omega \right)^2 +  \left( \partial_2 \omega \right)^2 +  \csc^2 \theta \left( \partial_3 \omega \right)^2 \right) \\
\cal{B}_\omega &= r^3 \left( \partial_0 \omega \right)^2 +  r e^\nu \left( r^2 e^{-\lambda}\left( \partial_1 \omega \right)^2 -  \left( \partial_2 \omega \right)^2 -  \csc^2 \theta \left( \partial_3 \omega \right)^2 \right)\\
\cal{C}_\omega &= r^3 \left( \partial_0 \omega \right)^2 +  r e^\nu \left( -r^2 e^{-\lambda}\left( \partial_1 \omega \right)^2 +  \left( \partial_2 \omega \right)^2 -  \csc^2 \theta \left( \partial_3 \omega \right)^2 \right) \\
\cal{D}_\omega &= r^3 \left( \partial_0 \omega \right)^2 +  r e^\nu \left( -r^2 e^{-\lambda}\left( \partial_1 \omega \right)^2 -  \left( \partial_2 \omega \right)^2 +  \csc^2 \theta \left( \partial_3 \omega \right)^2 \right)\\
\end{align*}
Conservation laws and equations of motion: 
\begin{align*}
\nabla_\mu {\calG}^{\mu0} & =   \frac{\alpha ' e^{-2\lambda -\nu}}{8 r^2} \left\{ \left(e^{\lambda }-1\right)\left(-2\partial_{11} \nu- \left(\partial_{1} \nu\right)^2\right)+\left(e^{\lambda }-3\right) \partial_1 \lambda  \partial_{1} \nu \right\} \partial_0 \phi \\
&+ \frac{\alpha ' e^{-2\lambda -\nu}}{8 r^2} \left\{\left(e^{\lambda }-1\right)\left(2 \partial_{00} \lambda - \partial_{0} \lambda \partial_{0} \nu \right)+\left(e^{\lambda }+1\right) \left(\partial_{0} \lambda\right)^2\right\}e^{\lambda-\nu } \partial_0 \phi  \\
\nabla_\mu {\calG}^{\mu1} & = \frac{\alpha ' e^{-3 \lambda} }{8 r^2}\left\{ \left(e^{\lambda }-1\right)\left(2   \partial_{11} \nu+ \left(\partial_{1} \nu\right)^2\right)-\left(e^{\lambda }-3\right)  \partial_{1} \lambda \partial_{1} \nu\right\} \partial_1 \phi\\
&+ \frac{\alpha ' e^{-3 \lambda} }{8 r^2} \left\{\left(e^{\lambda }-1\right)\left(-2  \partial_{00} \lambda + \partial_0 \lambda \partial_0 \nu\right)-\left(e^{\lambda }+1\right) \left(\partial_0 \lambda\right)^2\right\}e^{\lambda-\nu } \partial_1 \phi   \\
\nabla_\mu {\calG}^{\mu2} & =  \frac{\alpha ' e^{-2 \lambda } }{8 r^4}\left\{\left(e^{\lambda }-1\right)\left(2    \partial_{11} \nu+  \left(\partial_{1} \nu\right)^2\right)-\left(e^{\lambda }-3\right)  \partial_{1} \lambda \partial_{1} \nu\right\} \partial_2 \phi\\
&+\frac{\alpha ' e^{-2 \lambda } }{8 r^4}\left\{ \left(e^{\lambda }-1\right)\left(-2  \partial_{00} \lambda + \partial_{0} \lambda \partial_{0} \nu\right)-\left(e^{\lambda }+1\right) \left(\partial_{0} \lambda\right)^2\right\}e^{\lambda-\nu }  \partial_2 \phi \\
\nabla_\mu {\calG}^{\mu3} & =  \frac{\alpha ' \csc ^2\theta e^{-2 \lambda} }{8 r^4}\left\{\left(e^{\lambda }-1\right) \left(2   \partial_{11} \nu+  \left(\partial_{1} \nu\right)^2\right)-\left(e^{\lambda }-3\right)  \partial_{1} \lambda \partial_{1} \nu\right\} \partial_3 \phi\\
&+ \frac{\alpha ' \csc ^2\theta e^{-2 \lambda} }{8 r^4} \left\{\left(e^{\lambda }-1\right) \left(-2 \partial_{00} \lambda + \partial_{0} \lambda \partial_{0} \nu\right)-\left(e^{\lambda }+1\right) \left(\partial_{0} \lambda\right)^2\right\}e^{\lambda -\nu } \partial_3 \phi \\
\end{align*}
\begin{align*}
\nabla_\mu {\calT}^{\mu0} & =   \frac{e^{-\lambda -2 \nu } }{2 r^2}\left\{e^{2 \phi} \left[e^{\lambda +\nu }  \left(r^2 P \left[\partial_0 \lambda+2 \partial_0 \nu\right]+2 r^2 \partial_0 \rho+r^2 \rho \left[4 \partial_0 \phi+\partial_0 \lambda+2 \partial_0 \nu\right]\right)\right]\right\}\\
& +\frac{e^{-\lambda -2 \nu } }{2 r^2}\left\{e^{2 \phi} \left[e^{\lambda+\nu }  \left( -2 \partial_0 \varphi \left[\partial_{22} \varphi+\csc ^2\theta \left(\partial_{33} \varphi+2 \partial_3 \phi \partial_3 \varphi\right)+\partial_2 \varphi \left(2 \partial_2 \phi+\cot \theta\right)\right]\right)\right]\right\}\\
& +\frac{e^{-\lambda -2 \nu } }{2 r^2}\left\{e^{2 \phi} \left[e^{\lambda +\nu}   \left(2 \partial_0 \phi \left[\left(\partial_2 \varphi\right)^2+\csc ^2\theta \left(\partial_3 \varphi\right)^2\right]\right)+r^2 e^{\lambda } \partial_0 \varphi \left(2 \partial_{00} \varphi+\partial_0 \varphi \left[2 \partial_0 \phi+\partial_0 \lambda-\partial_0 \nu\right]\right)\right]\right\}\\
& +\frac{e^{-\lambda -2 \nu } }{2 r^2}\left\{e^{2 \phi} \left[r e^\nu  \left(2 r \partial_0 \phi \left(\partial_1 \varphi\right)^2-\partial_0 \varphi \left[2 r \partial_{11} \varphi+\partial_1 \varphi \left(r \left[4 \partial_1 \phi-\partial_1 \lambda+\partial_1 \nu\right]+4\right)\right]\right)\right]\right\}\\
& +\frac{e^{-\lambda -2 \nu } }{2 r^2}\left\{\partial_0 \phi \left[ r^2e^{\lambda } \left(2 \partial_{00} \phi+\left(\partial_0 \lambda-\partial_0 \nu\right) \partial_0 \phi\right)-2 e^{\lambda+\nu}  \left(\partial_{22} \phi+\csc ^2\theta \partial_{33} \phi+\cot \theta \partial_2 \phi\right)\right]\right\}\\
& +\frac{e^{-\lambda -2 \nu } }{2 r^2}\left\{\partial_0 \phi \left[-r e^\nu  \left(2 r \partial_{11} \phi+\left[-r \partial_1 \lambda+r \partial_1 \nu+4\right] \partial_1 \phi\right)\right]\right\}\\
 \nabla_\mu {\calT}^{\mu1} & =  \frac{e^{-2 (\lambda +\nu )}}{2 r^2} \left\{e^{2 \phi } \left[e^{\lambda +\nu } \left(2 e^\nu  \left[\partial_1 \phi \left(-\left(\partial_2 \varphi\right)^2-\csc ^2\theta \left(\partial_3 \varphi\right)^2+2 r^2 P\right)+r^2 \partial_1 P\right]\right)\right]\right\}\\
 &+ \frac{e^{-2 (\lambda +\nu )}}{2 r^2} \left\{e^{2 \phi } \left[e^{\lambda +\nu } \left(2 e^\nu  \left[\partial_1 \varphi \left(\partial_{22} \varphi +\csc ^2\theta \left[\partial_{33} \varphi +2 \partial_3 \phi \partial_3 \varphi\right]+\partial_2 \varphi \left[2 \partial_2 \phi+\cot \theta\right]\right)\right]\right)\right]\right\}\\
 &+ \frac{e^{-2 (\lambda +\nu )}}{2 r^2} \left\{e^{2 \phi } \left[e^{\lambda +\nu } \left(r^2 e^\nu  \partial_1 \nu (P+\rho )+r^2 \left[\partial_0 \varphi \left(2 \partial_1 \phi \partial_0 \varphi+\partial_1 \varphi \left[-4 \partial_0 \phi-\partial_0 \lambda+\partial_0 \nu\right]\right)-2 \partial_1 \varphi \partial_{00} \varphi \right]\right)\right]\right\}\\
  &+ \frac{e^{-2 (\lambda +\nu )}}{2 r^2} \left\{e^{2 \phi } \left[r e^{2 \nu } \partial_1 \varphi \left(2 r \partial_{11} \varphi +\partial_1 \varphi \left[r \left(2 \partial_1 \phi-\partial_1 \lambda+\partial_1 \nu\right)+4\right]\right)\right]    \right\}\\
 &+ \frac{e^{-2 (\lambda +\nu )}}{2 r^2} \left\{e^{\lambda+\nu}  \partial_1 \phi \left[ r^2\left(\partial_0 \nu-\partial_0 \lambda\right) \partial_0 \phi-2r^2 \partial_{00} \phi +2 e^\nu  \left(\partial_{22} \phi +\csc ^2\theta \partial_{33} \phi +\cot \theta \partial_2 \phi\right)   \right]\right\}\\
  &+ \frac{e^{-2 (\lambda +\nu )}}{2 r^2} \left\{e^\nu  \partial_1 \phi \left[r e^\nu  \left(2 r \partial_{11} \phi +\left[-r \partial_1 \lambda+r \partial_1 \nu+4\right] \partial_1 \phi\right)\right]\right\}\\
 \nabla_\mu {\calT}^{\mu2} & =  \frac{e^{-\lambda -\nu } }{2 r^4}\left\{e^{2 \phi } \left[4 r^2 e^{\lambda +\nu } P \partial_2 \phi+\partial_2 \varphi \left(r^2e^{\lambda } \left[\partial_0 \varphi \left(-4 \partial_0 \phi-\partial_0 \lambda+\partial_0 \nu\right)-2 \partial_{00} \varphi \right]\right)\right]\right\}\\
 &+\frac{e^{-\lambda -\nu } }{2 r^4}\left\{e^{2 \phi } \left[\partial_2 \varphi \left(2e^{\lambda+\nu }   \left[\partial_{22} \varphi +\csc ^2\theta \left(\partial_{33} \varphi +2 \partial_3 \phi \partial_3 \varphi\right)+\cot \theta \partial_2 \varphi\right]\right)\right]\right\}\\
&+\frac{e^{-\lambda -\nu } }{2 r^4}\left\{e^{2 \phi } \left[\partial_2 \varphi \left(r e^\nu  \left[2 r \partial_{11} \varphi +\partial_1 \varphi \left(r \left[4 \partial_1 \phi-\partial_1 \lambda+\partial_1 \nu\right]+4\right)\right]\right)\right]\right\}\\
&+\frac{e^{-\lambda -\nu } }{2 r^4}\left\{e^{2 \phi } \left[-2 \partial_2 \phi \left(r^2 e^\nu  \left(\partial_1 \varphi\right)^2-e^{\lambda } \left[r^2 \left(\partial_0 \varphi\right)^2+e^\nu  \left(\left(\partial_2 \varphi\right)^2-\csc ^2\theta \left(\partial_3 \varphi\right)^2\right)\right]\right)\right]\right\}\\
&+\frac{e^{-\lambda -\nu } }{2 r^4}\left\{\partial_2 \phi \left[r^2e^{\lambda } \left(\left[\partial_0 \nu-\partial_0 \lambda\right] \partial_0 \phi-2 \partial_{11} \phi \right)+2e^{\lambda+\nu }  \left(\partial_{22} \phi +\csc ^2\theta \partial_{33} \phi +\cot \theta \partial_2 \phi\right)\right]\right\}\\
&+\frac{e^{-\lambda -\nu } }{2 r^4}\left\{\partial_2 \phi \left[r e^\nu  \left(2 r \partial_{11} \phi +\left[-r \partial_1 \lambda+r \partial_1 \nu+4\right] \partial_1 \phi\right)\right]\right\} \\
\nabla_\mu {\calT}^{\mu3} & =   \frac{\csc ^2\theta e^{-\lambda -\nu } }{2 r^4}\left\{e^{2 \phi} \left[4 r^2 e^{\lambda +\nu } P \partial_3 \phi+\partial_3 \varphi \left(r^2e^{\lambda } \left[\partial_0 \varphi \left(-4 \partial_0 \phi-\partial_0 \lambda+\partial_0 \nu\right)-2 \partial_{00} \varphi\right]\right)\right]\right\}\\
&+\frac{\csc ^2\theta e^{-\lambda -\nu } }{2 r^4}\left\{e^{2 \phi} \left[\partial_3 \varphi \left(2e^{\lambda +\nu}  \left[\partial_{22} \varphi+\csc ^2\theta \partial_{33} \varphi+\partial_2 \varphi \left(2 \partial_2 \phi+\cot \theta\right)\right]\right)\right]\right\}\\
&+\frac{\csc ^2\theta e^{-\lambda -\nu } }{2 r^4}\left\{e^{2 \phi} \left[\partial_3 \varphi \left(r e^\nu  \left[2 r \partial_{11} \varphi+\partial_1 \varphi \left(r \left[4 \partial_1 \phi-\partial_1 \lambda+\partial_1 \nu\right]+4\right)\right]\right)\right]\right\}\\
&+\frac{\csc ^2\theta e^{-\lambda -\nu } }{2 r^4}\left\{e^{2 \phi} \left[-2 \partial_3 \phi \left( e^{\lambda+\nu }  \left[\left(\partial_2 \varphi\right)^2-\csc ^2\theta \left(\partial_3 \varphi\right)^2\right]-r^2e^{\lambda } \left(\partial_0 \varphi\right)^2+r^2 e^\nu  \left(\partial_1 \varphi\right)^2\right)\right]\right\}\\
&+\frac{\csc ^2\theta e^{-\lambda -\nu } }{2 r^4}\left\{\partial_3 \phi \left[r^2e^{\lambda } \left(\left[\partial_0 \nu-\partial_0 \lambda\right] \partial_0 \phi-2 \partial_{00} \phi\right)+2e^{\lambda +\nu} \left(\partial_{22} \phi+\csc ^2\theta \partial_{33} \phi+\cot \theta \partial_2 \phi\right)\right]\right\}\\
&+\frac{\csc ^2\theta e^{-\lambda -\nu } }{2 r^4}\left\{\partial_3 \phi \left[r e^\nu  \left(2 r \partial_{11} \phi+\left[-r \partial_1 \lambda+r \partial_1 \nu+4\right] \partial_1 \phi\right)\right]\right\} \\
\end{align*}
\begin{align*}
    \nabla^2\phi &= \frac{1}{2 r^2}\left\{r^2 e^{-\nu } \left[\left(\partial_0 \nu-\partial_0 \lambda\right) \partial_0 \phi -2 \partial_{00} \phi \right]+e^{-\lambda } r \left[2 r \partial_{11} \phi +\left(-r \partial_1 \lambda+r \partial_1 \nu+4\right) \partial_1 \phi \right]\right\}\\
    &+\frac{1}{2 r^2}\left\{2 \left[\partial_{22} \phi +\csc ^2\theta \partial_{33} \phi +\cot \theta \partial_2 \phi \right]\right\}\\
    e^{2\phi }(\partial\varphi )^2 - \frac{\alpha'}{8}e^{-\phi}\mathcal{X}_4 - 8\pi T^{\text{mat}}  &= \frac{1}{4 r^2}\left\{\alpha ' \left[2 \left(e^{\lambda }-1\right) e^\nu \partial_{11} \nu+\left(e^{\lambda }-1\right) e^\nu \left(\partial_{1} \nu\right)^2-\left(e^{\lambda }-3\right) e^\nu \partial_{1} \lambda \partial_{1} \nu\right] e^{-\phi -2 \lambda -\nu}\right\}\\
    &+\frac{1}{4 r^2}\left\{\alpha ' \left[ -2 \left[e^{\lambda }-1\right]e^{\lambda } \partial_{00} \lambda-\left(e^{\lambda }+1\right)e^{\lambda } \left(\partial_{0} \lambda\right)^2+\left[e^{\lambda }-1\right] e^{\lambda }\partial_{0} \lambda \partial_{0} \nu\right] e^{-\phi -2 \lambda -\nu}\right\}\\
    &+e^{2 \phi } \left\{\frac{\left(\partial_2 \varphi \right)^2+\csc ^2\theta \left(\partial_3 \varphi \right)^2}{r^2}+e^{-\lambda } \left(\partial_1 \varphi \right)^2-e^{-\nu} \left(\partial_0 \varphi \right)^2\right\}-8 \pi  (3 P-\rho ) e^{2 \phi } \\
    \nabla_{\mu}(e^{2\phi }\nabla^{\mu}\varphi ) &=   \frac{1}{2 r^2}\left\{e^{2 \phi -\lambda -\nu } \left[r^2e^{\lambda } \left(\partial_0 \varphi  \left[-4 \partial_0 \phi -\partial_0 \lambda +\partial_0 \nu \right]-2 \partial_{00} \varphi \right)\right]\right\}\\
    &+\frac{1}{2 r^2}\left\{e^{2 \phi -\lambda -\nu } \left[2e^{\lambda +\nu }   \left(\partial_{22} \varphi +\csc ^2\theta \left[\partial_{33} \varphi +2 \partial_3 \phi  \partial_3 \varphi \right]+\partial_2 \varphi  \left[2 \partial_2 \phi +\cot \theta\right]\right)\right]\right\}\\
     &+\frac{1}{2 r^2}\left\{e^{2 \phi -\lambda -\nu } \left[r e^\nu  \left(2 r \partial_{11} \varphi +\partial_1 \varphi  \left[r \left(4 \partial_1 \phi -\partial_1 \lambda +\partial_1 \nu \right)+4\right]\right)\right]\right\} \\
    -\frac{\alpha'}{8}R_{\mu\nu\rho\sigma}\Tilde{R}^{\mu\nu\rho\sigma} &= 0
\end{align*}

Metric perturbation equivalence:
\allowdisplaybreaks
\begin{align*}
    \Box h_{\mu\nu} &= \frac{\alpha'}{8}\bigg(D_{\mu\nu}^{(\phi)} + 2C_{\mu\nu}\bigg) \\
    D_{00}^{(\phi)} &= \frac{e^{-2 \lambda }}{r^3} \left\{2 \left(e^{\lambda }-1\right) r \left(e^{\lambda } \partial_0 \lambda \partial_0 \phi-2 e^\nu  \partial_{11} \phi\right)+2 e^\nu  \partial_1 \lambda \left(-e^{\lambda } \left[\partial_{22} \phi+\csc ^2\theta \partial_{33} \phi-r \partial_1 \phi+\cot \theta \partial_2 \phi\right]-3 r \partial_1 \phi\right)\right\}\\
    D_{11}^{(\phi)} &= \frac{2 \left\{\left(e^{\lambda }-1\right) r e^{-\nu} \left(\partial_0 \nu \partial_0 \phi-2 \partial_{00} \phi\right)+\partial_1 \nu \left[-\partial_{22} \phi+e^{-\lambda } \left(e^{\lambda }-3\right) r \partial_1 \phi+\csc ^2\theta \left(-\partial_{33} \phi\right)-\cot \theta \partial_2 \phi\right]\right\}}{r^3} \\
    D_{22}^{(\phi)} &= e^{-\lambda -\nu} \left\{r \partial_0 \lambda \left(4 \partial_{01} \phi-\partial_1 \nu \partial_0 \phi\right)-e^\nu \left(2 \partial_{11} \nu-\partial_1 \lambda \partial_1 \nu+\left(\partial_1 \nu\right)^2\right) \left(\csc ^2\theta \partial_{33} \phi+\cot \theta \partial_2 \phi\right)\right\}\\
    &+e^{-\lambda -\nu} \left\{-r \left[\partial_0 \lambda \left(\partial_0 \lambda+\partial_0 \nu\right)-2 \partial_{00} \lambda\right] \partial_1 \phi+r \partial_1 \lambda \left(\partial_0 \nu \partial_0 \phi-2 \partial_{00} \phi\right)\right\}\\
    &-e^{-2 \lambda } r \left\{\left(2 \partial_{11} \nu-3 \partial_1 \lambda \partial_1 \nu+\left(\partial_1 \nu\right)^2\right) \partial_1 \phi+2 \partial_1 \nu \partial_{11} \phi\right\}-e^{-\nu} \left\{\partial_0 \lambda \left(\partial_0 \nu-\partial_0 \lambda\right)-2 \partial_{00} \lambda\right\} \left\{\csc ^2\theta \partial_{33} \phi+\cot \theta \partial_2 \phi\right\} \\
    D_{33}^{(\phi)} &= \sin ^2\theta e^{-2 \lambda -\nu} \left\{e^{\lambda } \left[r \partial_0 \lambda \left(4 \partial_{01} \phi-\partial_1 \nu \partial_0 \phi\right)-e^\nu \left(2 \partial_{11} \nu-\partial_1 \lambda \partial_1 \nu+\left(\partial_1 \nu\right)^2\right) \partial_{22} \phi\right]\right\}\\
    &+\sin ^2\theta e^{-2 \lambda -\nu} \left\{e^{\lambda } \left[-r \left(\partial_0 \lambda \left[\partial_0 \lambda+\partial_0 \nu\right]-2 \partial_{00} \lambda\right) \partial_1 \phi+r \partial_1 \lambda \left(\partial_0 \nu \partial_0 \phi-2 \partial_{00} \phi\right)\right]\right\}\\
    &+\sin ^2\theta e^{-2 \lambda -\nu} \left\{-re^\nu \left[\left(2 \partial_{11} \nu-3 \partial_1 \lambda \partial_1 \nu+\left(\partial_1 \nu\right)^2\right) \partial_1 \phi+2 \partial_1 \nu \partial_{11} \phi\right]-e^{2 \lambda } \left[\partial_0 \lambda \left(\partial_0 \nu-\partial_0 \lambda\right)-2 \partial_{00} \lambda\right] \partial_{22} \phi\right\} \\
    D_{01} &= \frac{e^{-\lambda } \left\{2 \left(e^{\lambda }-1\right) r \left(\partial_1 \nu \partial_0 \phi-2 \partial_{01} \phi\right)+2 \partial_0 \lambda \left[-e^{\lambda } \left(\partial_{22} \phi+\csc ^2\theta \partial_{33} \phi-r \partial_1 \phi+\cot \theta \partial_2 \phi\right)-3 r \partial_1 \phi\right]\right\}}{r^3} = D_{10}\\
    D_{02} &= -\frac{2 e^{-\lambda } \left\{\partial_0 \lambda \left(\partial_2 \phi-r \partial_{12} \phi\right)+r \partial_1 \lambda \partial_{02} \phi\right\}}{r^2} = D_{20}\\
    D_{03} &= -\frac{2 e^{-\lambda } \left\{\partial_0 \lambda \left(\partial_3 \phi-r \partial_{13} \phi\right)+r \partial_1 \lambda \partial_{03} \phi\right\}}{r^2} = D_{30}\\
    D_{12} &= \frac{2 \left\{e^{-\lambda } \partial_1 \nu \left(r \partial_{12} \phi-\partial_2 \phi\right)-r e^{-\nu} \partial_0 \lambda \partial_{02} \phi\right\}}{r^2} = D_{21}\\
    D_{13} &= \frac{2 \left\{e^{-\lambda } \partial_1 \nu \left(r \partial_{13} \phi-\partial_3 \phi\right)-r e^{-\nu} \partial_0 \lambda \partial_{03} \phi\right\}}{r^2} = D_{31}\\
    D_{23} &= -e^{-\lambda -\nu} \left\{e^\nu \left[2 \partial_{11} \nu-\partial_1 \lambda \partial_1 \nu+\left(\partial_1 \nu\right)^2\right]+e^{\lambda } \left[\partial_0 \lambda \left(\partial_0 \nu-\partial_0 \lambda\right)-2 \partial_{00} \lambda\right]\right\} \left(\cot \theta \partial_3 \phi-\partial_{23} \phi\right) = D_{32}\\
    &\\
    &\\
    &\\
    &\\
    C_{02} &= -\frac{\csc \theta e^{\nu -\frac{3 (\lambda +\nu )}{2}} }{8 r^3}\left\{ \left[r^2e^{\lambda } \left[2 \partial_{00} \lambda+\left(\partial_0 \lambda\right)^2-\partial_0 \lambda \partial_0 \nu\right]+4e^{\lambda +\nu}\right] \left(\partial_3 \varphi-r \partial_{13} \varphi\right) \right\}\\
    &-\frac{\csc \theta e^{\nu -\frac{3 (\lambda +\nu )}{2}} }{8 r^3}\left\{\left[-e^\nu  \left( 2 r^2 \partial_{11} \nu+r\left(r \partial_{1} \nu-2\right) \left(\partial_{1} \nu-\partial_{1} \lambda\right)+4\right)\right] \left(\partial_3 \varphi-r \partial_{13} \varphi\right)\right\}\\
    &-\frac{\csc \theta e^{\nu -\frac{3 (\lambda +\nu )}{2}} }{8 r^3}\left\{\left[r^2e^{\lambda } \left(-2 r \partial_{001} \lambda+r \partial_0 \lambda \left[\partial_{01} \nu-2 \partial_{01} \lambda\right]+\partial_0 \nu \left[r \partial_{01} \lambda+\left(3-r \partial_{1} \nu\right) \partial_0 \lambda\right]\right)\right] \partial_3 \varphi\right\}\\
    &-\frac{\csc \theta e^{\nu -\frac{3 (\lambda +\nu )}{2}} }{8 r^3}\left\{\left[r^2e^{\lambda } \left(2 \left[r \partial_{1} \nu-3\right] \partial_{00} \lambda+\left(r \partial_{1} \nu-3\right) \left(\partial_0 \lambda\right)^2\right)-4e^{\lambda +\nu}\right] \partial_3 \varphi\right\}\\
    &-\frac{\csc \theta e^{\nu -\frac{3 (\lambda +\nu )}{2}} }{8 r^3}\left\{\left[4e^\nu -re^\nu  \left(\partial_{1} \nu \left[ r^2 \left[\partial_{11} \lambda-2 \partial_{11} \nu\right]-r^2 \left(\partial_{1} \lambda\right)^2+r\partial_{1} \lambda+4\right]\right)\right] \partial_3 \varphi\right\}\\
    &-\frac{\csc \theta e^{\nu -\frac{3 (\lambda +\nu )}{2}} }{8 r^3}\left\{\left[-r^2e^\nu  \left(3 r \partial_{1} \lambda \partial_{11} \nu-2 \left[r \partial_{111} \nu+\partial_{11} \lambda+2 \partial_{11} \nu\right]+2 \left(\partial_{1} \lambda\right)^2\right)-r^2e^\nu  \left(r \partial_{1} \lambda-3\right) \left(\partial_{1} \nu\right)^2\right] \partial_3 \varphi\right\}= C_{20} \\
    C_{03} & = -\frac{\sin \theta e^{\nu -\frac{3 (\lambda +\nu )}{2}} }{8 r^3}\left\{\left[e^{\lambda } \left(r^2 \left[\partial_0 \lambda \left(\partial_0 \nu-\partial_0 \lambda\right)-2 \partial_{00} \lambda\right]-4 e^\nu \right)\right] \left(\partial_2 \varphi-r \partial_{12} \varphi\right)\right\}\\
    &-\frac{\sin \theta e^{\nu -\frac{3 (\lambda +\nu )}{2}} }{8 r^3}\left\{\left[e^\nu  \left(2 r^2 \partial_{11} \nu+r\left[r \partial_{1} \nu-2\right] \left[\partial_{1} \nu-\partial_{1} \lambda\right]+4\right)\right] \left(\partial_2 \varphi-r \partial_{12} \varphi\right)\right\}\\
    &-\frac{\sin \theta e^{\nu -\frac{3 (\lambda +\nu )}{2}} }{8 r^3}\left\{ \left[r^2e^{\lambda } \left(2 r \partial_{001} \lambda-r \partial_0 \lambda \left[\partial_{01} \nu-2 \partial_{01} \lambda\right]+\partial_0 \nu \left[\left(r \partial_{1} \nu-3\right) \partial_0 \lambda-r \partial_{01} \lambda\right]\right)\right] \partial_2 \varphi\right\}\\
    &-\frac{\sin \theta e^{\nu -\frac{3 (\lambda +\nu )}{2}} }{8 r^3}\left\{ \left[r^2e^{\lambda } \left(-2 \left[r \partial_{1} \nu-3\right] \partial_{00} \lambda+\left[3-r \partial_{1} \nu\right] \left(\partial_0 \lambda\right)^2\right)+4e^{\lambda +\nu}\right] \partial_2 \varphi\right\}\\
    &-\frac{\sin \theta e^{\nu -\frac{3 (\lambda +\nu )}{2}} }{8 r^3}\left\{\left[re^\nu  \left(\partial_{1} \nu \left[r \left(r \left[\partial_{11} \lambda-2 \partial_{11} \nu\right]-r \left(\partial_{1} \lambda\right)^2+\partial_{1} \lambda\right)+4\right]+r \left[3 r \partial_{1} \lambda \partial_{11} \nu\right]\right)\right] \partial_2 \varphi\right\}\\
    &-\frac{\sin \theta e^{\nu -\frac{3 (\lambda +\nu )}{2}} }{8 r^3}\left\{\left[ re^\nu \left(r \left[-2 \left(r \partial_{111} \nu+\partial_{11} \lambda+2 \partial_{11} \nu\right)+2 \left(\partial_{1} \lambda\right)^2\right]+r \left[r \partial_{1} \lambda-3\right] \left(\partial_{1} \nu\right)^2\right)-4e^\nu\right] \partial_2 \varphi\right\} = C_{30} \\
    C_{12} & = \frac{\csc \theta e^{\lambda -\frac{3 (\lambda +\nu )}{2}}}{8 r^2} \left\{\left[r^2e^{\lambda } \left(2 \partial_{00} \lambda+\left(\partial_0 \lambda\right)^2-\partial_0 \lambda \partial_0 \nu\right)+4 e^{\lambda +\nu}-re^\nu \left(2 r \partial_{11} \nu+\left[r \partial_{1} \nu-2\right] \left[\partial_{1} \nu-\partial_{1} \lambda\right]\right)-4e^\nu\right] \partial_{03} \varphi\right\}\\
   &+\frac{\csc \theta e^{\lambda -\frac{3 (\lambda +\nu )}{2}}}{8 r^2} \left\{ \left[re^\nu  \left(-2 r \partial_{011} \nu+\left[r \partial_{1} \lambda-2 r \partial_{1} \nu+2\right] \partial_{01} \nu+\left[r \partial_{1} \nu-2\right] \partial_{01} \lambda\right)\right] \partial_3 \varphi\right\}\\
   &+\frac{\csc \theta e^{\lambda -\frac{3 (\lambda +\nu )}{2}}}{8 r^2} \left\{ \left[e^\nu  \left( r\partial_0 \lambda \left[2 r \partial_{11} \nu+\left(r \partial_{1} \nu-2\right) \left(\partial_{1} \nu-\partial_{1} \lambda\right)\right]+4\partial_0 \lambda\right)\right] \partial_3 \varphi\right\}\\
   &+\frac{\csc \theta e^{\lambda -\frac{3 (\lambda +\nu )}{2}}}{8 r^2} \left\{\left[r^2e^{\lambda }  \left(2 \partial_{000} \lambda-\partial_0 \lambda \left[\partial_{00} \nu+\partial_0 \nu \left(\partial_0 \lambda-\partial_0 \nu\right)\right]+\partial_{00} \lambda \left[2 \partial_0 \lambda-3 \partial_0 \nu\right]\right)\right] \partial_3 \varphi\right\} = C_{21} \\
    C_{13} &= \frac{\sin \theta e^{\lambda -\frac{3 (\lambda +\nu )}{2}} }{8 r^2}\left\{\left[ r^2e^{\lambda } \left(\partial_0 \lambda \left[\partial_0 \nu-\partial_0 \lambda\right]-2 \partial_{00} \lambda\right)-4e^{\lambda+\nu } +  re^\nu \left(2 r \partial_{11} \nu+\left[r \partial_{1} \nu-2\right] \left[\partial_{1} \nu-\partial_{1} \lambda\right]\right)+4e^\nu\right] \partial_{02} \varphi\right\}\\
    &+\frac{\sin \theta e^{\lambda -\frac{3 (\lambda +\nu )}{2}} }{8 r^2}\left\{\left[r^2e^{\lambda }  \left(-2 \partial_{000} \lambda+\partial_0 \lambda \left[\partial_{00} \nu+\partial_0 \nu \left(\partial_0 \lambda-\partial_0 \nu\right)\right]+\partial_{00} \lambda \left[3 \partial_0 \nu-2 \partial_0 \lambda\right]\right)\right] \partial_2 \varphi\right\}\\
    &+\frac{\sin \theta e^{\lambda -\frac{3 (\lambda +\nu )}{2}} }{8 r^2}\left\{\left[ -re^\nu \left(-2 r \partial_{011} \nu+\left[r \partial_{1} \lambda-2 r \partial_{1} \nu+2\right] \partial_{01} \nu+\left[r \partial_{1} \nu-2\right] \partial_{01} \lambda\right)\right] \partial_2 \varphi\right\}\\
    &+\frac{\sin \theta e^{\lambda -\frac{3 (\lambda +\nu )}{2}} }{8 r^2}\left\{\left[-e^\nu  \left(r\partial_0 \lambda \left[2 r \partial_{11} \nu+\left(r \partial_{1} \nu-2\right) \left(\partial_{1} \nu-\partial_{1} \lambda\right)\right]+4\partial_0 \lambda\right)\right] \partial_2 \varphi\right\} = C_{31} \\  
\end{align*}
\newpage
%%%%%%%%%%%%%%%%%%%%%%%%%%%%%%%%%%%%%%%%%%%%%%%%%%%%%%%%%%%%%%%%%%%%%%%%%%%%%%%%%%%%%%%%%%%%%%%%%%%%%%%%%%%%%%%%%%%%%

\section{Schwarzschild metric}\label{sec:Schwarzschild}
Field equations: 
\begin{align*}
{\cal{G}}_{00} &=  \frac{1}{4 r^6} r_s \alpha ' (r-r_s)  \left[2 r (r_s-r) \partial_{11} \phi  +(2 r-3 r_s) \partial_1 \phi  +\partial_{22} \phi  + \csc^2\theta \partial_{33} \phi  + \cot\theta \partial_2 \phi  \right] \\
{\cal{G}}_{11} & =  \frac{1}{4 r^4}  r_s\alpha ' \frac{1}{(r-r_s)^2}\left\{-2 r^3 \partial_{00} \phi  -(r-r_s) \left[(2 r-3 r_s) \partial_1 \phi  +\partial_{22} \phi  + \csc^2\theta \partial_{33} \phi  + \cot\theta \partial_2 \phi  \right]\right\} \\
{\cal{G}}_{22} & =  \frac{1}{4r^3} r_s \alpha ' \left[r (r_s-r) \partial_{11} \phi  +(2 r-3 r_s) \partial_1 \phi  +2  \csc^2\theta \partial_{33} \phi  +2  \cot\theta \partial_2 \phi  +\left(\frac{r^3}{r-r_s}\right) \partial_{00} \phi  \right] \\
{\cal{G}}_{33} & =  \frac{1}{4r^3} r_s \alpha ' \sin^2\theta \left[r (r_s-r) \partial_{11} \phi  +(2 r-3 r_s) \partial_{1} \phi  +2 \partial_{22} \phi  +\left( \frac{r^3 }{r-r_s}\right) \partial_{00} \phi \right] \\
{\cal{G}}_{01}&= \frac{1}{4 r^4} r_s \alpha ' \left\{ \left( \frac{r_s  }{r-r_s}\right) \partial_0 \phi  -2 r \partial_{01}\phi  \right\} ={\cal{G}}_{10}\\
{\cal{G}}_{02}&=    \frac{1}{16r^5} r_s \alpha ' \left[4 r^2 \partial_{02}\phi  +3  \csc\theta (r_s-r)  \left(\partial_3 \varphi  -r \partial_{13}\varphi  \right)\right] = {\cal{G}}_{20}\\
{\cal{G}}_{03}&=  \frac{1}{16r^5} r_s \alpha ' \left[4 r^2 \partial_{03}\phi  -3 \sin\theta (r_s-r)   \left(\partial_2 \varphi  -r \partial_{12}\varphi  \right)\right] = {\cal{G}}_{30}  \\
{\cal{G}}_{12 }&=  \frac{1}{16r^4} r_s \alpha ' \left\{r \left[  3 \csc\theta  \partial_{03}\varphi \left(\frac{ r   }{r-r_s} \right) +4 \partial_{12}\phi  \right]-4 \partial_2 \phi  \right\} = {\cal{G}}_{21} \\
{\cal{G}}_{13 }&=   \frac{1}{16 r^4}r_s \alpha ' \left\{r \left[  3 \sin\theta  \partial_{02}\varphi\left(\frac{ r   }{r_s-r}\right) +4 \partial_{13}\phi  \right]-4 \partial_3 \phi  \right\} = {\cal{G}}_{31} \\
{\cal{G}}_{23 }&=   \frac{1}{2r^3} r_s \alpha ' \left[ \cot\theta \partial_3 \phi  -\partial_{23}\phi  \right] = {\cal{G}}_{32 } \\
{\calT}_{00} & =  \frac{1}{2r^3} \left\{e^{2 \phi  } \left[  {\cal{A}}_\varphi \right] + {\cal{A}}_\phi +2 r^2(r-r_s) \rho   e^{2 \phi  }\right\} \\
{\cal{T}}_{11} & =  \frac{1}{2r(r- r_s)^2} \left\{ e^{2 \phi  } \left[ {\cal{B}}_\varphi \right] + \left[ {\cal{B}}_\phi \right] + 2r^2 \left(r-r_s\right) P e^{2 \phi}\right\} \\
{\cal{T}}_{22 } & =   \frac{1}{2 \left( r-r_s \right)} \left\{e^{2 \phi }\left[ {\cal{C}_{\varphi}} \right] + \left[ {\cal{C}_{\phi}} \right] +2r^2 (r - r_s) P e^{2\phi}\right\}  \\
{\cal{T}}_{33 } & =  \frac{\sin^2{\theta}}{2(r-r_s)} \left\{ e^{2 \phi  } \left[ {\cal{D}_\varphi} \right] + {\cal{D}_\phi} + 2r^2(r-r_s)Pe^{2\phi}  \right\} \\
{\cal{T}}_{\mu \nu}&={ e^{2 \phi  } \partial_\mu \varphi  \partial_\nu \varphi +\partial_\mu \phi  \partial_\nu \phi = \cal{T}}_{\nu \mu} , \, \, \,  \mu \neq \nu \\
\end{align*}
where for $\omega =  \left\{ \varphi,\phi \right\}$, 
\begin{align*} 
{\cal{A}}_\omega &= r^3 \left(\partial_0 \omega \right)^2 +( r - r_s) \left[ r (r-r_s) \left(\partial_1 \omega \right)^2+\left(\partial_2 \omega \right)^2+\csc^2 \theta  \left(\partial_3 \omega \right)^2 \right]  \\
\cal{B}_\omega &=  r^3 \left(\partial_0 \omega \right)^2 +( r - r_s) \left[ r (r-r_s) \left(\partial_1 \omega \right)^2-\left(\partial_2 \omega \right)^2-\csc^2 \theta  \left(\partial_3 \omega \right)^2 \right]  \\
\cal{C}_\omega &=  r^3 \left(\partial_0 \omega \right)^2 +( r - r_s) \left[ -r (r-r_s) \left(\partial_1 \omega \right)^2+\left(\partial_2 \omega \right)^2-\csc^2 \theta  \left(\partial_3 \omega \right)^2 \right]  \\
\cal{D}_\omega &=  r^3 \left(\partial_0 \omega \right)^2 +( r - r_s) \left[ -r (r-r_s) \left(\partial_1 \omega \right)^2-\left(\partial_2 \omega \right)^2+\csc^2 \theta  \left(\partial_3 \omega \right)^2 \right]  \\
\end{align*}
\newpage
Conservation laws and equations of motion: 
\begin{align*}
\nabla_\mu {\calG}^{\mu0}  = & 3 r_s^2 \alpha ' \left[ \frac{ \partial_0 \phi}{4 r^5 (r-r_s)} \right] \\
\nabla_\mu {\calG}^{\mu1}  = & 3 r_s^2 \alpha ' \left[  \frac{ (r_s-r) \partial_1 \phi }{4 r^7} \right] \\
\nabla_\mu {\calG}^{\mu2}  = & - 3 r_s^2 \alpha ' \left[ \frac{ \partial_2 \phi }{4 r^8} \right] \\
\nabla_\mu {\calG}^{\mu3}  = & - 3 r_s^2 \alpha '   \left[ \frac{ \csc ^2\theta  \partial_3 \phi }{4 r^8} \right] \\
\nabla_\mu {\calT}^{\mu0}  = & \frac{e^{2 \phi  } \left\{ r^3  \partial_0 \varphi   \partial_{00} \varphi +r^3  \partial_{0} \phi  \left( \partial_{1} \varphi \right)^2+r^3  \partial_{0} \phi  \left( \partial_0 \varphi \right)^2-2 r^3  \partial_{1} \phi   \partial_{1} \varphi   \partial_0 \varphi -r r_s^2  \partial_{11} \varphi   \partial_0 \varphi \right\} }{r (r-r_s)^2}\\
+& \frac{ e^{2 \phi  } \left\{ 2 r^2 (r-r_s) \rho    \partial_{0} \phi -2 r^2 r_s  \partial_{0} \phi  \left( \partial_{1} \varphi \right)^2+4 r^2 r_s  \partial_{1} \phi   \partial_{1} \varphi   \partial_0 \varphi +2 r^2 r_s  \partial_{11} \varphi   \partial_0 \varphi -2 r^2  \partial_{1} \varphi   \partial_0 \varphi  \right\}}{r (r-r_s)^2}\\
+& \frac{ e^{2 \phi  } \left\{-r^3  \partial_{11} \varphi   \partial_0 \varphi+  r r_s^2  \partial_{0} \phi  \left( \partial_{1} \varphi \right)^2-2 r r_s^2  \partial_{1} \phi   \partial_{1} \varphi   \partial_0 \varphi -r_s^2  \partial_{1} \varphi   \partial_0 \varphi  + r_s  \partial_0 \varphi   \partial_{22} \varphi -r_s  \partial_{0} \phi  \left( \partial_{2} \varphi \right)^2 \right\} }{r (r-r_s)^2}\\
+& \frac{ e^{2 \phi  } \left\{ \csc ^2 \theta  (r-r_s) \left[ \partial_{0} \phi  \left( \partial_{3} \varphi \right)^2- \partial_0 \varphi  \left( \partial_{33} \varphi +2  \partial_{3} \phi   \partial_{3} \varphi \right)\right]+2 r_s  \partial_{2} \phi   \partial_0 \varphi   \partial_{2} \varphi +  3 r r_s  \partial_{1} \varphi   \partial_0 \varphi \right\}}{r (r-r_s)^2} \\
+& \frac{ e^{2 \phi  } \left\{ \cot \theta (r_s-r)  \partial_0 \varphi   \partial_{2} \varphi -r  \partial_0 \varphi   \partial_{22} \varphi + r  \partial_{0} \phi  \left( \partial_{2} \varphi \right)^2-2 r  \partial_{2} \phi   \partial_0 \varphi   \partial_{2} \varphi +r^2 (r-r_s) \partial_0 \rho \right\}}{r (r-r_s)^2}\\
+& \frac{   \partial_{0} \phi  \left\{ r^3  \partial_{00} \phi -r (r-r_s)^2  \partial_{11} \phi -(r-r_s) \left[(2 r-r_s)  \partial_{1} \phi + \partial_{22} \phi + \csc ^2 \theta   \partial_{33} \phi +\cot \theta  \partial_{2} \phi \right] \right\}  }{r (r-r_s)^2} \\
\nabla_\mu {\calT}^{\mu1}  = & \frac{e^{2 \phi }\left\{r P\left[4 r\left(r-r_s\right) \partial_1 \phi +r_s\right]+2 \partial_1 \phi \left[r^3 \left(\partial_0 \varphi \right)^2-\left(r-r_s\right) \left( \left(\partial_2 \varphi \right)^2+  \csc^2\theta \left(\partial_3 \varphi \right)^2\right)\right]\right\}}{2 r^3}\\ 
+ &\frac{e^{2 \phi }\left\{2 \partial_1 \varphi \left[-r^3 \partial_{00} \varphi -2 r^3 \partial_0 \varphi  \partial_0 \phi +r \left(r-r_s\right)^2 \partial_{11} \varphi -r_s \partial_{22} \varphi +\csc ^2 \theta \left(r-r_s\right) \left(\partial_{33} \varphi +2 \partial_3 \varphi  \partial_3 \phi \right)\right]\right\}}{2 r^3}  \\
+ & \frac{e^{2 \phi }\left\{2\partial_1 \varphi \left[ \cot\theta \left(r-r_s\right)\partial_2 \varphi  +r \partial_{22} \varphi+2 r \partial_2 \phi \partial_2 \varphi  +\left(r-r_s\right) \partial_1 \varphi \left(r\left(r-r_s\right) \partial_1 \phi+2 r -r_s\right)\right]\right\}}{2 r^3} \\
+ & \frac{2 \partial_1 \phi \left\{\left(r-r_s\right) \left[  \cot\theta  \partial_2 \phi + r\left(r-r_s\right) \partial_{11} \phi+\left(2 r-r_s\right) \partial_1 \phi+\partial_{22} \phi + \csc^2\theta \partial_{33} \phi   \right]-r^3 \partial_{00} \phi \right\}}{2 r^3}\\
 + & \frac{e^{2 \phi }\left\{-2 r_s \partial_1 \varphi \partial_2 \varphi \partial_2 \phi+  r r_s \rho\right\}}{2 r^3} \\
\nabla_\mu {\calT}^{\mu2}  = & \frac{e^{2 \phi  } \left\{2 r^2 P  (r-r_s)  \partial_{2} \phi + \partial_{2} \phi  \left[r^3 \left( \partial_0 \varphi \right)^2-r (r-r_s)^2 \left( \partial_{1} \varphi \right)^2+(r-r_s) \left(\left( \partial_{2} \varphi \right)^2- \csc ^2 \theta  \left( \partial_{3} \varphi \right)^2\right)\right] \right\}  }{r^4 (r-r_s)} \\
+ & \frac{e^{2 \phi  } \left\{ \partial_{2}\varphi  \left[-r^3  \partial_{00} \varphi -2 r^3  \partial_{0} \phi   \partial_0 \varphi +r (r-r_s)^2  \partial_{11} \varphi -r_s  \partial_{22} \varphi + \csc ^2 \theta  (r-r_s) \left( \partial_{33} \varphi +2  \partial_{3} \phi   \partial_{3} \varphi \right) \right] \right\} }{r^4 (r-r_s)} \\
+& \frac{e^{2 \phi  } \left\{ \partial_{2} \varphi  \left[
(r-r_s)  \partial_{1} \varphi  \left(2 r (r-r_s)  \partial_{1} \phi +2 r-r_s\right)+\cot \theta (r-r_s)  \partial_{2} \varphi +r  \partial_{22} \varphi \right]\right\} }{r^4 (r-r_s)} \\
+&  \frac{  \partial_{2} \phi  \left\{(r-r_s) \left[r (r-r_s)  \partial_{11} \phi +(2 r-r_s)  \partial_{1} \phi + \partial_{22} \phi + \csc ^2 \theta   \partial_{33} \phi +\cot \theta  \partial_{2} \phi \right]-r^3  \partial_{00} \phi \right\} }{r^4 (r-r_s)} \\
\nabla_\mu {\calT}^{\mu3}  = &  \frac{ \csc ^2 \theta  \left\{e^{2 \phi  } \left[2 r^2 P  (r-r_s)  \partial_{3} \phi + \partial_{3} \phi  \left[r^3 \left( \partial_0 \varphi \right)^2+(r-r_s) \left(r (r_s-r) \left( \partial_{1} \varphi \right)^2-\left( \partial_{2} \varphi \right)^2\right) \right] \right] \right\}}{r^4 (r-r_s)}
\\
+& \frac{ \csc ^2 \theta  \left\{e^{2 \phi  } \left[ \partial_{3} \varphi  \left(-r^3  \partial_{00} \varphi -2 r^3  \partial_{0} \phi   \partial_0 \varphi +r (r-r_s)^2  \partial_{11} \varphi -r_s  \partial_{22} \varphi + \csc ^2 \theta  (r-r_s)  \partial_{33} \varphi \right) \right] \right\} }{r^4 (r-r_s)}
\\
+&\frac{ \csc ^2 \theta  \left\{e^{2 \phi  } \left( \partial_3 \varphi \left[ (r-r_s)  \partial_{1} \varphi  \left(2 r (r-r_s)  \partial_{1} \phi +2 r-r_s\right)-2 r_s  \partial_{2} \phi   \partial_{2} \varphi -r_s \cot \theta  \partial_{2} \varphi +r  \partial_{22} \varphi \right] \right )\right\} }{r^4 (r-r_s)}
\\
+&  \frac{ \csc ^2 \theta  \left\{ \partial_{3} \phi  \left[(r-r_s) \left(r (r-r_s)  \partial_{11} \phi +(2 r-r_s)  \partial_{1} \phi + \partial_{22} \phi + \csc ^2 \theta   \partial_{33} \phi +\cot \theta  \partial_{2} \phi \right)-r^3  \partial_{00} \phi \right] \right\}}{r^4 (r-r_s)} \\
+&  \frac{ \csc ^2 \theta  \left\{e^{2 \phi  } \left[ r \cot \theta  \partial_{2} \varphi \partial_3 \varphi  +  \left(r-r_s\right) \csc ^2 \theta \partial_3 \phi \left( \partial_{3} \varphi \right)^2 +2 r \partial_3 \varphi \partial_{2} \phi   \partial_{2} \varphi \right] \right\}}{r^4 (r-r_s)}. \\
\end{align*}
\begin{align*}
\nabla^2\phi &= \frac{r (r-r_s) \partial_{11} \phi  +(2 r-r_s) \partial_1 \phi  +\partial_{22} \phi  + \csc^2\theta \partial_{33} \phi  + \cot\theta \partial_2 \phi  }{r^2}-\frac{r \partial_{00} \phi  }{r-r_s} \\
e^{2\phi }(\partial\varphi )^2 - \frac{\alpha'}{8}e^{-\phi}\mathcal{X}_4 - 8\pi T^{\text{mat}}  &= \frac{e^{2 \phi  } \left[\frac{r^3 \left(\partial_0 \varphi  \right)^2}{r_s-r}+r (r-r_s) \left(\partial_1 \varphi  \right)^2+\left(\partial_2 \varphi  \right)^2+ \csc^2\theta \left(\partial_3 \varphi  \right)^2\right]}{r^2} \nonumber\\
&-8 \pi  [3 P(t)-\rho (t) ]  e^{2 \phi  }-\frac{3 r_s^2 \alpha ' e^{-\phi  }}{2 r^6}\\
\nabla_{\mu}(e^{2\phi }\nabla^{\mu}\varphi ) &= \frac{e^{2 \phi  } \left\{-r^3 \partial_{00} \varphi  +2 r^3 \partial_1 \phi   \partial_1 \varphi  -2 r^3 \partial_0 \phi   \partial_0 \varphi  +r r_s^2 \partial_{11} \varphi  -4 r^2 r_s \partial_1 \phi   \partial_1 \varphi  -2 r^2 r_s \partial_{11} \varphi  +2 r^2 \partial_1 \varphi \right\} }{r^2 (r-r_s)} \\
 &+ \frac{e^{2 \phi  } \left\{ r^3 \partial_{11} \varphi  +2 r r_s^2 \partial_1 \phi   \partial_1 \varphi  +r_s^2 \partial_1 \varphi  -r_s \partial_{22} \varphi  + \csc^2\theta (r-r_s) \left(\partial_{33} \varphi  +2 \partial_3 \phi   \partial_3 \varphi  \right)\right\}}{r^2 (r - r_s)} \\
&+ \frac{e^{2 \phi  } \left\{ -2 r_s \partial_2 \phi   \partial_2 \varphi - 3 r r_s \partial_1 \varphi  + \cot\theta (r-r_s) \partial_2 \varphi  +r \partial_{22} \varphi  +2 r \partial_2 \phi   \partial_2 \varphi  \right\}}{r^2 (r-r_s)}\\
-\frac{\alpha'}{8}R_{\mu\nu\rho\sigma}\Tilde{R}^{\mu\nu\rho\sigma} &= 0
\end{align*}

Metric perturbation equivalence:

\begin{align*}
\Box h_{\mu\nu} &= \frac{\alpha'}{8}\bigg(D_{\mu\nu}^{(\phi)} + 2C_{\mu\nu}\bigg) \\
D_{00}^{(\phi)} &= \frac{2 r_s (r-r_s) \left\{2 r (r_s-r) \partial_{11} \phi+(2 r-3 r_s) \partial_{1} \phi+\partial_{22} \phi+\csc ^2\theta \partial_{33} \phi+\cot \theta \partial_{2} \phi\right\}}{r^6} \\
D_{11}^{(\phi)} &= \frac{-4 r^3 r_s \partial_{00} \phi-2 r_s (r-r_s) \left\{(2 r-3 r_s) \partial_{1} \phi+\partial_{22} \phi+\csc ^2\theta \partial_{33} \phi+\cot \theta \partial_{2} \phi\right\}}{r^4 (r-r_s)^2} \\
D_{22}^{(\phi)} &= \frac{2 r_s \left\{r^3 \partial_{00} \phi+(r_s-r) \left[r (r-r_s) \partial_{11} \phi+(3 r_s-2 r) \partial_{1} \phi-2 \csc ^2\theta \partial_{33} \phi-2 \cot \theta \partial_{2} \phi\right]\right\}}{r^3 (r-r_s)} \\
D_{33}^{(\phi)} &= 2 r_s \sin ^2\theta \left\{\frac{r (r_s-r) \partial_{11} \phi+(2 r-3 r_s) \partial_{1} \phi+2 \partial_{22} \phi}{r^3}+\frac{\partial_{00} \phi}{r-r_s}\right\} \\
D_{01}^{(\phi)} &= \frac{2 r_s \left\{r_s \partial_{0} \phi-2 r (r-r_s)\partial_{01} \phi\right\}}{r^4(r-r_s)} = D_{10}^{(\phi)}\\
D_{02}^{(\phi)} &= \frac{2 r_s \partial_{02} \phi}{r^3} = D_{20}^{(\phi)}\\
D_{03}^{(\phi)} &= \frac{2 r_s \partial_{03} \phi}{r^3} = D_{30}^{(\phi)}\\
D_{12}^{(\phi)} &= \frac{2 r_s \left\{r \partial_{12} \phi-\partial_{2} \phi\right\}}{r^4} = D_{21}^{(\phi)}\\
D_{13}^{(\phi)} &= \frac{2 r_s \left\{r \partial_{13} \phi-\partial_{3} \phi\right\}}{r^4} = D_{31}^{(\phi)}\\
D_{23}^{(\phi)} &= \frac{4 r_s \left\{\cot \theta \partial_{3} \phi-\partial_{23} \phi\right\}}{r^3} = D_{32}^{(\phi)}\\
C_{02} &= \frac{3 r_s \csc \theta (r-r_s) \left\{r \partial_{13} \varphi -\partial_3 \varphi \right\}}{2 r^5} = C_{20} \\
C_{03} &= -\frac{3 r_s \sin \theta (r-r_s) \left\{r \partial_{12}\varphi -\partial_{2} \varphi \right\}}{2 r^5} = C_{30} \\
C_{12} &= \frac{3 r_s \csc \theta \partial_{03}\varphi }{2 r^2 (r-r_s)} = C_{21} \\
C_{13} &= \frac{3 r_s \sin \theta \partial_{02} \varphi}{2 r^2 (r_s-r)} = C_{31} \\
\end{align*}

\newpage 
%%%%%%%%%%%%%%%%%%%%%%%%%%%%%%%%%%%%%%%%%%%%%%%%%%%%%%%%%%%%%%%%%%%%%%%%%%%%%%%%%%%%%%%%%%%%%%%%%%%%%%%%%%%%%%%%%%%%%%

\section{Perturbed Minkowski metric}\label{sec:PertMinkowski}
\begin{align*}
{\cal{G}}_{00} & =   \frac{3\left(\partial_0 \lambda \right)^2}{2(1-2 \lambda )^2}  \left\{2-\frac{\alpha ' \partial_0 \lambda  \partial_0 \phi  }{(2 \gamma +1) (1 -2\lambda)}\right\}\\
{\cal{G}}_{ii} & = \frac{1}{2 (2 \gamma +1)^3 (1-2 \lambda )^2}\left\{\alpha ' \partial_0 \lambda (2 \gamma +1) \left[(2 \lambda -1) \partial_0 \lambda \partial_{00} \phi   -2 \left((1-2 \lambda ) \partial_{00} \lambda +\left(\partial_0 \lambda\right)^2\right) \partial_0 \phi  \right]\right\} \\
&+ \frac{1}{2 (2 \gamma +1)^3 (1-2 \lambda )^2}\left\{3\alpha ' \partial_0 \lambda   (1-2 \lambda ) \partial_0 \gamma \partial_0 \lambda  \partial_0 \phi \right\}\\
&+\frac{1}{2 (2 \gamma +1)^3 (1-2 \lambda )^2}\left\{-2 (2 \gamma +1) (1-2 \lambda ) \left[2 (1-2 \lambda ) \partial_0 \gamma  \partial_0 \lambda -(2 \gamma +1) \left(2 (1-2 \lambda ) \partial_{00} \lambda +\left(\partial_0 \lambda \right)^2\right)\right]\right\} \\
{\cal{T}}_{00} & =  \frac{1}{2} \left\{e^{2 \phi  } \left(\partial_0 \varphi  \right)^2+\left(\partial_0 \phi  \right)^2+2 (2 \gamma +1) \rho   e^{2 \phi  }\right\} \\ 
{\cal{T}}_{ii} & = \frac{(2 \lambda -1) \left\{-e^{2 \phi  } \left(\left(\partial_0 \varphi  \right)^2+(4 \gamma +2) P \right)-\left(\partial_0 \phi  \right)^2\right\}}{4 \gamma +2}, \, \, \,  i = 1,2,3.
\end{align*}
\begin{align*} 
\nabla_\mu {\calT}^{\mu0} & =   \frac{1}{2 (2 \gamma +1)^3}\left\{2 (2 \gamma +1) e^{2 \phi} \left[2 P \partial_0 \gamma +\partial_0 \varphi \left(\partial_{00} \varphi+\partial_0 \phi \partial_0 \varphi\right)\right]\right\}\\
& + \frac{1}{2 (2 \gamma +1)^3}\left\{2 (2 \gamma +1) \left[e^{2 \phi} \left(2 \rho  \left[(2 \gamma +1) \partial_0 \phi+2 \partial_0 \gamma \right]+(2 \gamma +1)\partial_0 \rho\right)+\partial_0 \phi \partial_{00} \phi \right]\right\}\\
&+\frac{1}{2 (2 \gamma +1)^3}\left\{\frac{3 (2 \gamma +1) \partial_0 \lambda  \left[e^{2 \phi } \left(\left(\partial_0 \varphi \right)^2+(4 \gamma +2) P\right)+\left(\partial_0 \phi \right)^2\right]}{2 \lambda -1}\right\}\\
&+\frac{1}{2 (2 \gamma +1)^3}\left\{\frac{\left[(2 \lambda -1) \partial_0 \gamma +3 (2 \gamma +1) \partial_0 \lambda \right] \left[e^{2 \phi } \left(\left(\partial_0 \varphi \right)^2+(4 \gamma +2) \rho \right)+\left(\partial_0 \phi \right)^2\right]}{2 \lambda -1}\right\}\\
&+\frac{1}{2 (2 \gamma +1)^3}\left\{-3 \partial_0 \gamma  \left[e^{2 \phi } \left(\left(\partial_0 \varphi \right)^2+(4 \gamma +2) \rho \right)+\left(\partial_0 \phi \right)^2\right]\right\}\\
\end{align*}
\begin{align*}
\nabla^2\phi &= \frac{(2 \gamma +1) \left\{\frac{3 \partial_0 \lambda  \partial_0 \phi  }{1-2 \lambda }-\partial_{00} \phi  \right\}+\partial_0 \gamma  \partial_0 \phi  }{(2 \gamma +1)^2}\\
&\\
e^{2\phi }(\partial\varphi )^2 - \frac{\alpha'}{8}e^{-\phi}\mathcal{X}_4 - 8\pi T^{\text{mat}}  &= \frac{3 \alpha ' e^{-\phi  } \left(\partial_0 \lambda \right)^2 \left\{(2 \gamma +1) \left[(1-2 \lambda ) \partial_{00} \lambda +\left(\partial_0 \lambda \right)^2\right]+(2 \lambda -1) \partial_0 \gamma  \partial_0 \lambda \right\}}{(2 \gamma +1)^3 (1-2 \lambda )^4}\\
&-\frac{e^{2 \phi  } \left(\partial_0 \varphi  \right)^2}{2 \gamma +1}-\frac{8 \pi  e^{2 \phi  } \left\{(6 \gamma +3) P -(2 \gamma +1) \rho  \right\}}{2 \gamma +1} \\
&\\
\nabla_{\mu}(e^{2\phi }\nabla^{\mu}\varphi ) &=\frac{e^{2 \phi  } \left\{(2 \lambda -1) \partial_0 \gamma  \partial_0 \varphi  -(2 \gamma +1) \left[(2 \lambda -1) \left(\partial_{00} \varphi  +2 \partial_0 \phi   \partial_0 \varphi  \right)+3 \partial_0 \lambda  \partial_0 \varphi  \right]\right\}}{(2 \gamma +1)^2 (2 \lambda -1)}\\
-\frac{\alpha'}{8}R_{\mu\nu\rho\sigma}\Tilde{R}^{\mu\nu\rho\sigma} &=0 \\
\nabla_\mu {\calG}^{\mu0} & =   -\frac{3 \alpha ' \left(\partial_0 \lambda \right)^2 \partial_0 \phi \left\{(2 \gamma +1) \left[(1-2 \lambda ) \partial_{00} \lambda+\left(\partial_0 \lambda \right)^2\right]+(2 \lambda -1) \partial_0 \gamma \partial_0 \lambda \right\}}{2 (2 \gamma +1)^4 (1-2 \lambda )^4} 
\end{align*}

\end{widetext}
\twocolumngrid

\bibstyle{abbrv}

\bibliography{bibo}

\end{document}